\documentclass[prr,reprint,superscriptaddress,altaffillsymbol,amsmath,nofootinbib,longbibliography]{revtex4-1}

\pdfoutput=1

\usepackage{ifthen}
\usepackage{booktabs} 
\usepackage{natbib}

\usepackage{wasysym}
\usepackage{slashed} 
\usepackage{bbm}
\usepackage{mathrsfs}
\usepackage{amssymb}
\usepackage{dsfont}
\usepackage[export]{adjustbox}

\usepackage{color}

\definecolor{OurBlue}{rgb}{0.384,0.616,0.784}
\definecolor{OurRed}{rgb}{0.878,0.388,0.212}

\usepackage[
colorlinks=true, 
linkcolor=OurBlue,
citecolor=OurRed,
urlcolor=OurBlue,
]{hyperref}

\allowdisplaybreaks

\newcommand{\beqra}{\begin{eqnarray}}
\newcommand{\eeqra}{\end{eqnarray}}
\newcommand{\beq}{\begin{equation}}
\newcommand{\eeq}{\end{equation}}

\renewcommand{\epsilon}{\varepsilon}

\newcommand{\ket}[1]{\left| #1 \right\rangle}
\newcommand{\bra}[1]{\left\langle #1 \right|}
\newcommand{\braket}[2]{\left\langle #1 \big| #2 \right\rangle}

\usepackage[columnwise]{lineno} 

\begin{document}

\title{\boldmath Chiral Phonons as Dark Matter Detectors}

\author{Carl P. Romao}
\email[carl.romao@mat.ethz.ch]{}
\affiliation{Department of Materials, ETH Z\"urich, CH-8093 Z\"urich, Switzerland}

\author{Riccardo Catena}
\affiliation{Department of Physics, Chalmers University of Technology, SE-412 96 Göteborg, Sweden}

\author{Nicola A. Spaldin}
\affiliation{Department of Materials, ETH Z\"urich, CH-8093 Z\"urich, Switzerland}

\author{Marek Matas}
\email[marek.matas@mat.ethz.ch]{}
\affiliation{Department of Materials, ETH Z\"urich, CH-8093 Z\"urich, Switzerland}

\begin{abstract}


We propose a method for detecting single chiral phonons that will enable their use as dark-matter detectors. We suggest metal--organic frameworks (MOFs) as detector materials, as their flexibility yields low-energy chiral phonons with measurable magnetic moments, and their anisotropy leads to directional sensitivity, which mitigates background contamination. To demonstrate our proposal, we calculate the phononic structure of the MOF InF$_3$($4,4'$-bipyridine), and show that it has highly chiral acoustic phonons. Detection of such chiral phonons via their magnetic moments would dramatically lower the excitation energy threshold for dark matter detection to the energy of a single phonon. We show that single phonon detection in a MOF would extend detector reach ten or more orders of magnitude below current limits, enabling exploration of a multitude of as-yet-unprobed dark matter candidates.

\end{abstract}

\maketitle

\section{Introduction}

The nature of dark matter (DM) is a key open question in modern-day physics. Cosmological observations, such as the Bullet cluster collision~\cite{Clowe:2006eq}, suggest that it consists of a yet-unknown particle species~\cite{Bertone:2016nfn}. However, although its gravitational effects can be observed and tested across a wide range of scales and processes~\cite{Bertone:2016nfn, Planck:2018vyg, Persic:1995ru, Carney:2019pza}, its non-gravitational interaction with standard-model matter (SMM) has so far eluded detection.

The expected DM mass region of the theoretically well-motivated weakly interacting massive particle (WIMP) has guided the construction of detectors towards searching for nuclear recoil events causing excitations with energy thresholds $\mathcal{O} (\text{keV})$. The WIMP-motivated region of DM candidate masses has been probed with great precision using ton-scale detectors with high purity~\cite{LZ:2022ufs,PandaX-4T:2021bab,XENON:2018voc}. The absence of a convincing signal to date is shifting the focus of the community towards lighter DM candidates in the keV -- GeV mass range~\cite{Battaglieri:2017aum}.  

Such light DM particles would transfer less energy to a detector and therefore require lower energy detection mechanisms than the traditional nuclear recoil. As efforts to detect these DM masses are in an early stage, sub-gram-scale detectors operating for a short time place world-leading constraints such as the 4.3\,ng superconducting nanowire detector operating for 180\,h~\cite{Hochberg:2021yud, Hochberg:2016ajh, Hochberg:2016sqx}.

Phonons (quantized vibrational excitations) in crystals offer a promising avenue for detection of low-energy DM particles as their energies reach from fractions to hundreds of meV and their properties are highly controllable by choice of chemistry and microstructure \cite{qian2021phonon}. The formalism of phononic excitations caused by DM interactions is well-established~\cite{PhysRevLett.127.081804, Schutz:2016tid, PhysRevD.100.092007, Campbell-Deem:2022fqm, Trickle:2019nya, Trickle:2020oki, Griffin:2018bjn} with past works exploring conventional targets, such as LiF, NaF, Si, and Ge~\cite{PhysRevD.101.055004, NaF_dm} and evaluating how different interaction models affect the final excitation rate~\cite{Trickle:2020oki, PhysRevD.105.015010, Knapen:2017ekk}.

The detection of phonons is challenging, as, in addition to their small energies, they are also generated thermally at non-zero temperatures, are usually delocalized in bulk crystals, and generally decay on a ps timescale into the lowest energy (slow transverse acoustic) band \cite{maris1993anharmonic}. Proposed methods to detect phonons resulting from DM excitation include the CRESST~\cite{CRESST:2019jnq} and SuperCDMS~\cite{SuperCDMS:2022kse} experiments, which measure phonons that random-walk-like propagate into a thin layer of superconducting transition edge sensors (TES). This relies, however, on the generation of large numbers of phonons, since many are scattered during diffusion across the interface or thermalization in the TES. Consequently, it requires large deposited energies with current thresholds being $\mathcal{O}$(10)\,eV. While the detection of single phonons would vastly improve the energy resolution, the short lifetime of excited phonons renders their direct detection prohibitive. We address this challenge here with a system able to sense individual acoustic phonons, which are the stable decay products of the initial excitation. 

\begin{figure*}[ht]
    \centering
    \includegraphics[width=0.48\textwidth]{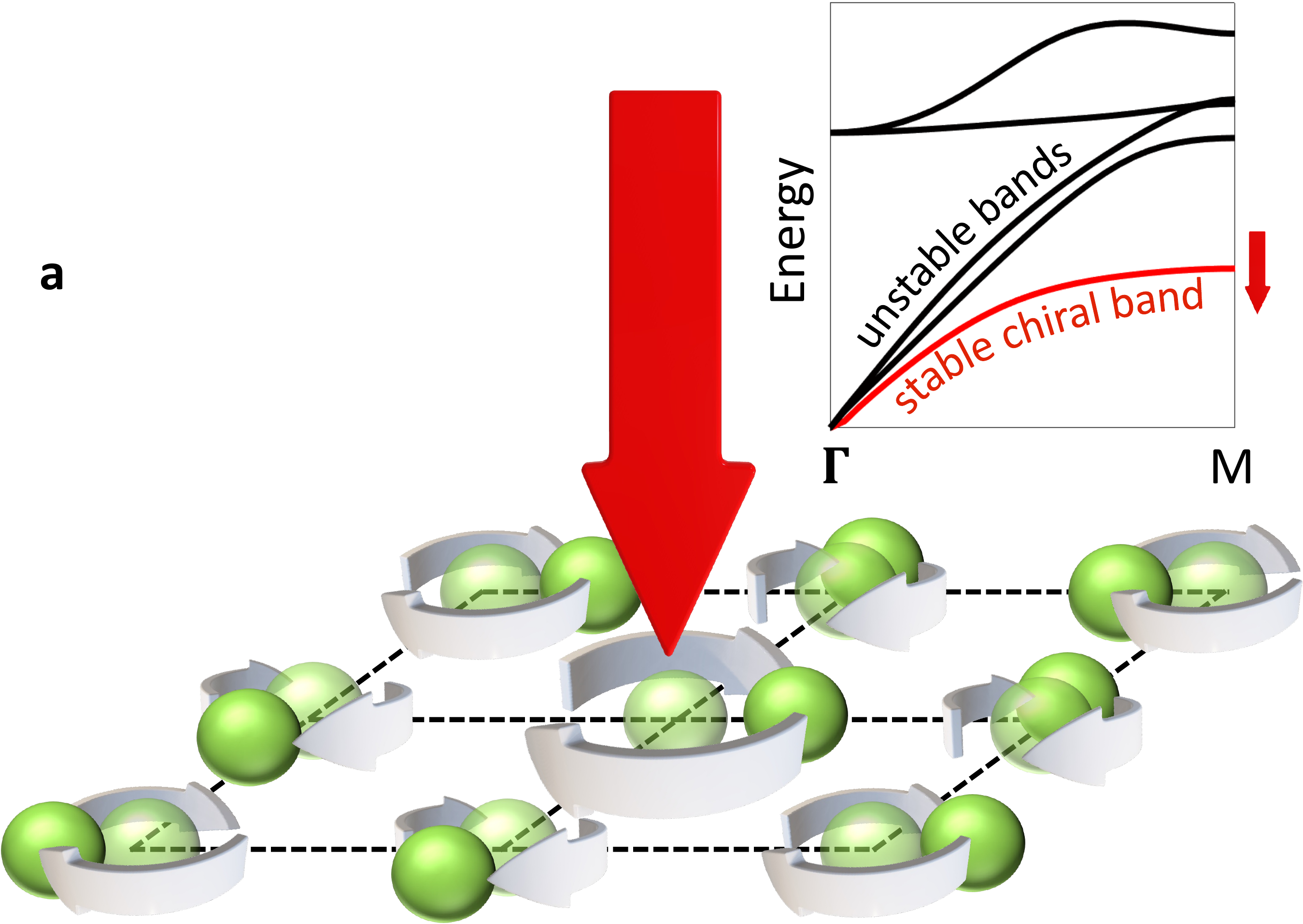}
    \includegraphics[width=0.48\textwidth]{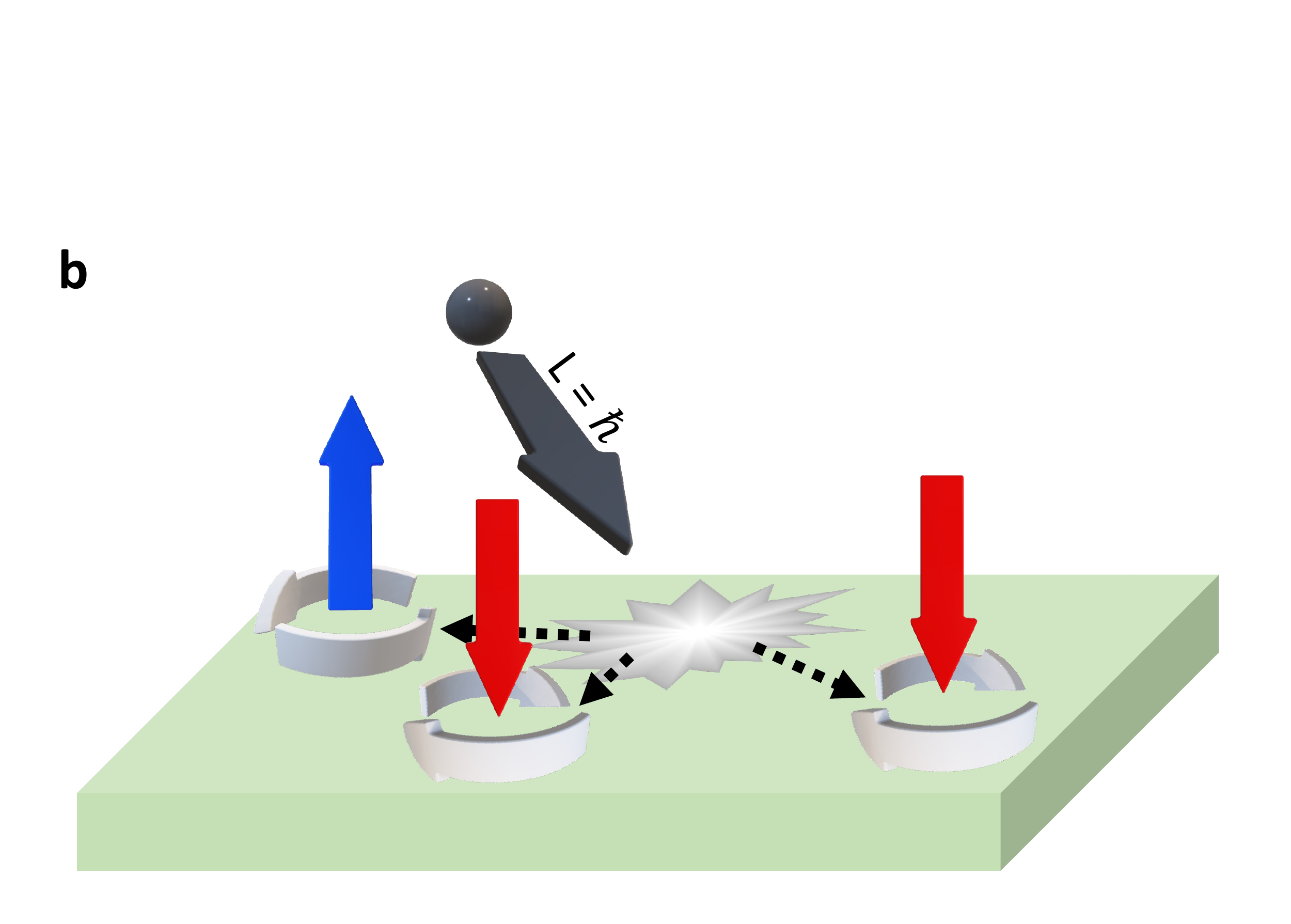}
    \includegraphics[width=0.48\textwidth]{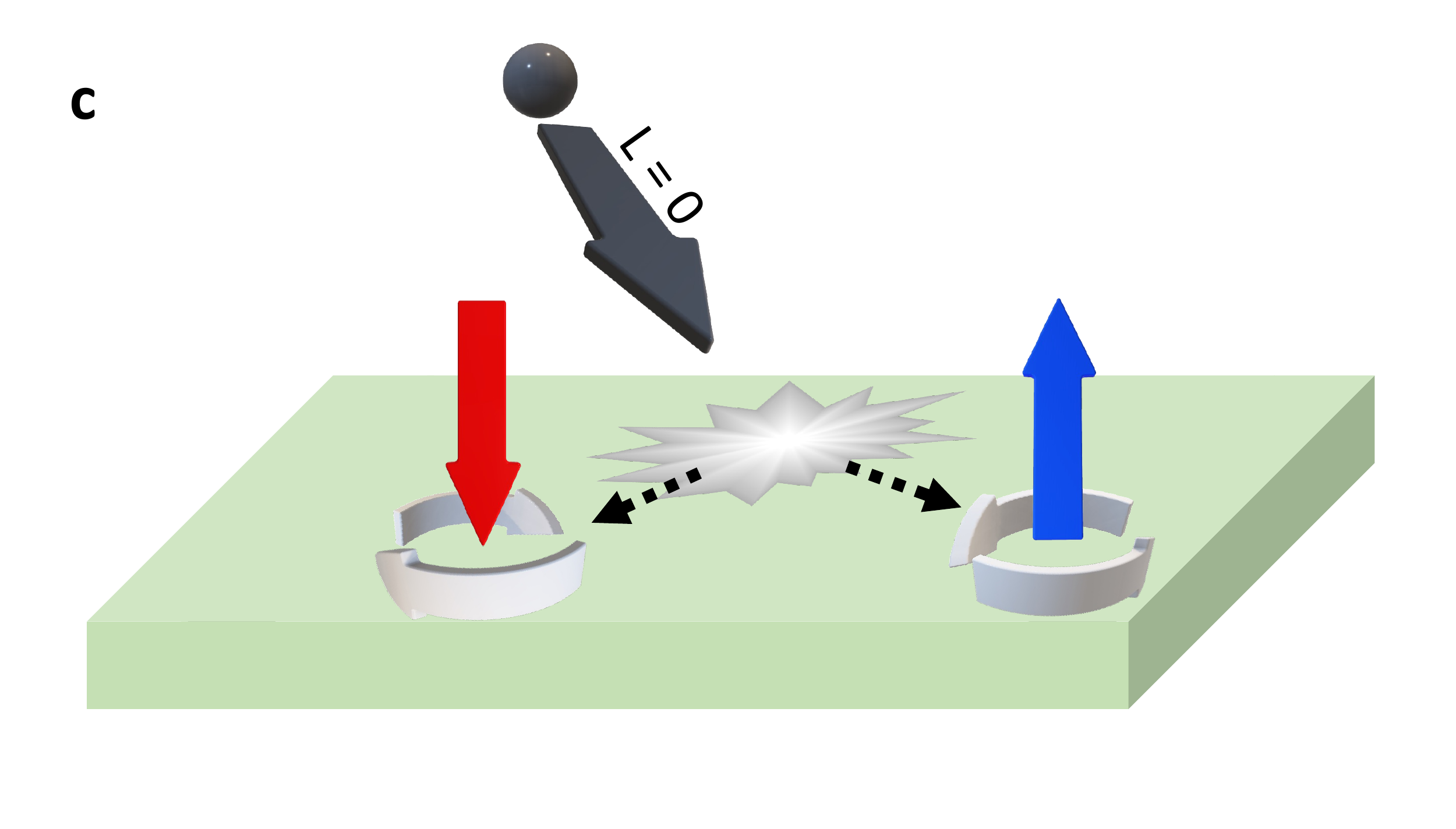}
    \includegraphics[width=0.48\textwidth]{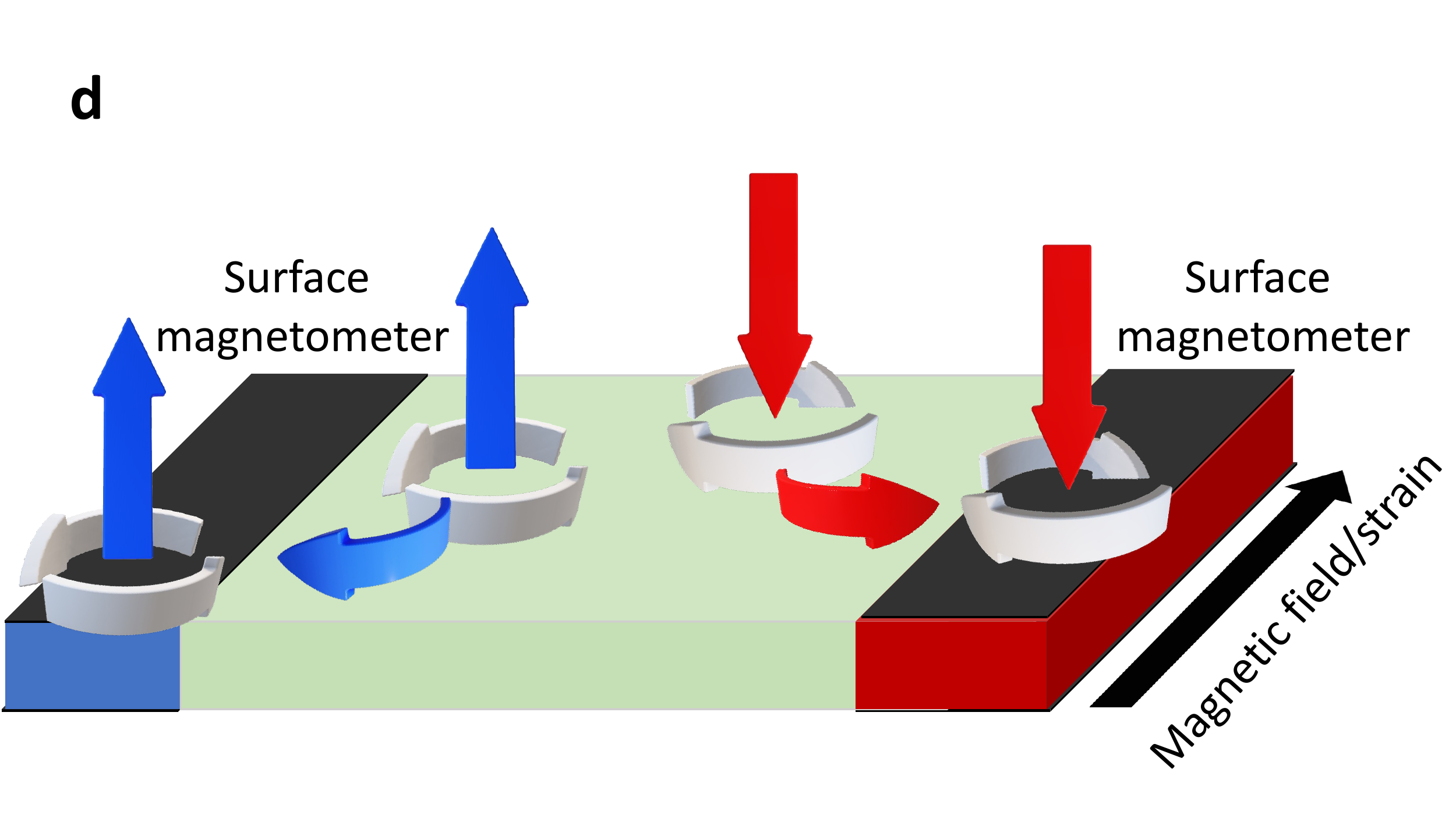}

  \caption{\textbf{a} The circular motion of atoms (green spheres) about their average positions (transparent green spheres) in a chiral phonon, generating a net angular momentum (arrow). The inset shows a cartoon phonon band structure highlighting the lowest energy band  (which can be chiral) in red; it contains phonons which are stable with respect to anharmonic decay \cite{maris1993anharmonic}. Energy and angular momentum will therefore be transferred from the higher energy bands, which are unstable with respect to decay, into this band. \textbf{b} Schematic view of a DM particle (black sphere) transferring energy and angular momentum into a material, creating several low-energy chiral phonons while conserving angular momentum. \textbf{c} A DM particle producing an achiral phonon which decays into a pair of phonons with opposite chirality. \textbf{d} Schematic view of a chiral phonon-based DM detector which uses the phonon thermal Hall effect to collect chiral phonons at the edges of the material in order to localize them for detection. A magnetic field or strain causes preferential transport of chiral phonons with opposite handedness in opposite directions and localization of the phonons near the surface magnetometers.}
  \label{fig-chiral-phonons}
\end{figure*}

We show that \textit{chiral} phonons \cite{zhang2015chiral} allow detection of excitations which produce only a single phonon, leading to $\mathcal{O} (\text{meV})$ detection energy thresholds. First, we demonstrate that established interactions between DM and SMM can create chiral phonons that carry angular momentum (Fig. \ref{fig-chiral-phonons} \textbf{a}). Then, we show that these chiral phonons can exist in the lowest energy band, where they are stable with respect to decay and can have a magnetic moment; the magnetic moments of such individual phonons have been found to reach detectable magnitudes \cite{schaack1976observation, BaydinPbTe, cheng2020large, hernandez2022chiral, juraschek2022giant, basini2022terahertz, luo2023large}. Next, we propose a detector architecture which uses the phonon Hall effect \cite{zhang2015chiral, park2020phonon} to collect the chiral phonons at the edges of a target (Fig. \ref{fig-chiral-phonons} \textbf{d}), allowing detection of one or more phonons and preserving directional information about the incoming DM particle. Finally, we demonstrate these principles with explicit numerical calculations for an example material, InF$_3$($4,4'$-bipyridine), and show that sensing of the magnetic moments of individual chiral phonons would allow the detection of scattering of DM particles down to $\mathcal{O} (\text{keV})$ masses (corresponding to $\mathcal{O} (\text{meV})$ energy thresholds), accessing vast regions of unexplored DM parameter space. 

\section{Chiral Phonons}

 Chiral phonons are quantized vibrational eigenmodes with an associated angular momentum, which arises from circular motions of atoms about their average positions (Fig.~\ref{fig-chiral-phonons} \textbf{a}). In materials that are both centrosymmetric and have time reversal symmetry, an inherent chirality of the phonon bands is forbidden. \cite{coh2023classification}. However, in noncentrosymmetric materials, phonons at low-symmetry points in reciprocal space break all mirror symmetries and are therefore chiral, and in magnetic materials any phonon can be circularly polarized \cite{coh2023classification}. Circularly polarized excitations can also be created from combinations of degenerate linear modes, \textit{e.g.} by circularly polarized light \cite{juraschek2017dynamical, juraschek2019orbital}.
 
The total angular momentum of the atoms ($\mathbf{L}$) can be expressed in terms of the atomic displacements ($\mathbf{u}_\alpha$) \cite{zhang2014angular}:
\begin{align}
    \label{eq-momentum}
    \mathbf{L} = \sum_\alpha \mathbf{u}_\alpha \times \dot{\mathbf{u}}_\alpha.
\end{align}
If we treat the atomic displacements within the harmonic model, we can write them in second quantization in terms of the creation ($\hat{a}^{\dagger}$) and annihilation ($\hat{a}$) operators of the phonon quasiparticle \cite{zhang2014angular}:
\begin{align}
    \label{eq-displacement}
    \mathbf{u}_\alpha = \sum_{n,\mathbf{q}}  e^{i(\mathbf{r}_\alpha \cdot \mathbf{q} - \omega_{n,\mathbf{q}} t)} \sqrt{\frac{\hbar}{2 \omega_{n,\mathbf{q}}N}} \hat{a}_{n,\mathbf{q}} \ket{\epsilon_{\alpha,n,\mathbf{q}}} + \mathrm{H.c.},
\end{align}
where $\ket{\epsilon_{\alpha,n,\mathbf{q}}}$ are the components of the phonon eigenvectors for atom $\alpha$ in band $n$ at wavevector $\mathbf{q}$, $\mathbf{r}$ is the position vector, $\omega$ is the phonon frequency, $N$ is the number of atoms participating in the phonon, and H.c. denotes the Hermitian conjugate. The eigenvectors are normalized such that $\sum_{\alpha=1}^{\alpha=N} \braket{\epsilon_{\alpha,n,\mathbf{q}}} {\epsilon_{\alpha,n,\mathbf{q}}} = 1$. In this notation, the creation operator for a right-circularly polarized phonon relative to $z$, $\hat{a}_{\mathrm{R}}^\dagger$, is:
\begin{align}
    \label{eq-createcircle}
    \hat{a}_{\mathrm{R},n,\mathbf{q}}^\dagger = \frac{1}{\sqrt{2}} (\hat{a}_{y,n,\mathbf{q}}^{\dagger} - i \hat{a}_{x,n,\mathbf{q}}^{\dagger}).
\end{align}

The $z$ component of the angular momentum operator is \cite{riseborough2010quantum}:
\begin{align}
    \label{eq-lhat}
    \hat{L}_z = i \hbar \sum_{n,\mathbf{q}} (\hat{a}_{x,n,\mathbf{q}}^{\dagger} \hat{a}_{y,n,\mathbf{q}} - \hat{a}_{y,n,\mathbf{q}}^{\dagger} \hat{a}_{x,n,\mathbf{q}}). 
\end{align}
A fully circularly polarized phonon has angular momentum $\hbar$ \cite{riseborough2010quantum}, as the commutator of these operators shows:
\begin{align}
    \label{eq-commutation}
    [\hat{L}_z, \hat{a}_{\mathrm{R},n,\mathbf{q}}^\dagger] = \hbar \hat{a}_{\mathrm{R},n,\mathbf{q}}^\dagger.
\end{align}
The $z$ component of the circular polarization operator ($\hat{S}_z = \ket{\mathrm{R}_z}\bra{\mathrm{R}_z} - \ket{\mathrm{\ell}_z}\bra{\mathrm{\ell}_z},$ where $\ket{\mathrm{R}_z}$ and $\ket{\mathrm{\ell}_z}$ are basis vectors for pure right- and left-handed rotations relative to $z$) is therefore related to $\hat{L}_z$ by $\hat{L}_z = \hbar\hat{S}_z$. Consequently, $\mathbf{L}$ is the sum of the phonon angular momenta $\mathbf{L}_{n,\mathbf{q}}$ and can be determined from the phonon eigenvectors \cite{zhang2015chiral}: 

\begin{align}
    \label{eq-pam}
    \mathbf{L} = \sum_{n,\mathbf{q}} \mathbf{L}_{n,\mathbf{q}} = \sum_{\alpha,n,\mathbf{q}} \mathbf{L}_{\alpha,n,\mathbf{q}} = \hbar \sum_{\alpha,n,\mathbf{q}} \bra{\epsilon_{\alpha,n,\mathbf{q}}} \hat{\bf S} \ket{\epsilon_{\alpha,n,\mathbf{q}}}.
\end{align}

 Since a phonon with a given energy and wavevector has a specific combination of angular and linear momentum  \cite{riseborough2010quantum, zhang2014angular}, the detection of a chiral phonon imparts information regarding the direction in which the exciting particle was traveling. This is a crucial feature of a DM detector since it allows separation of signal and noise by reorientation of the detector relative to the DM wind~\cite{Boyd:2022tcn}. 

Chiral phonons generate magnetic fields in materials, with the simplest model assuming a field from the circular motion of ionic charges following Ampère's law \cite{juraschek2017dynamical, juraschek2019orbital}. This mechanism yields magnetic moments on the order of the nuclear magneton ($\mu_\mathrm{n}$), which is difficult to detect using magnetometry. However, experimental investigations of phonon magnetism have found evidence of much larger, readily detectable, moments (on the order of $\mu_\mathrm{B}$, the Bohr magneton) in CsF$_3$ ($\mu_\mathrm{ph} = 7 \mu_\mathrm{B}$) \cite{luo2023large}, Cd$_3$As$_2$ ($\mu_\mathrm{ph} = 2.7 \mu_\mathrm{B}$) \cite{cheng2020large}, Pb$_{0.76}$Sn$_{0.24}$Te ($\mu_\mathrm{ph} = 1.5 \mu_\mathrm{B}$) \cite{hernandez2022chiral}, CsCl$_3$ ($\mu_\mathrm{ph} = 1.4 \mu_\mathrm{B}$) \cite{schaack1976observation, juraschek2022giant}, PbTe ($\mu_\mathrm{ph} = 0.6 \mu_\mathrm{B}$) \cite{BaydinPbTe}, and SrTiO$_3$ ($\mu_\mathrm{ph} = 0.1 \mu_\mathrm{B}$) \cite{basini2022terahertz}. 

While not fully understood, these giant phonon magnetic moments are believed to arise from interactions between the ionic and electronic angular momenta \cite{geilhufe2023electron}, \textit{e.g.} by coupling to the crystal field splitting of $4f$ paramagnets \cite{juraschek2022giant, luo2023large}, or to topological electronic states \cite{ren2021phonon, cheng2020large}. Such electronic contributions to the phonon magnetic moment can be expressed in terms of anomalously large effective charges ($\mathbf{Z}_\alpha^\star$) of the ions \cite{ren2021phonon}, such that the phonon magnetic moment is:

\begin{align} 
    \label{eq-magmo}
    \boldsymbol{\mu}_{\text{ph}} = \sum_{\alpha} \mathbf{L}_{\mathrm{ph},\alpha}  \frac{\mathbf{Z}_\alpha^\star}{2m_\alpha}
\end{align}
where $\mathbf{L}_{\mathrm{ph},\alpha}$ are the angular momenta of each atom participating in the phonon and $m_\alpha$ are the atomic masses \cite{quartz}.

\section{DM--Phonon Coupling}

There are two ways in which DM particles can interact with their SMM counterparts. They can scatter (exchanging momentum and kinetic energy) or be absorbed (depositing their entire four-momentum in the interaction). For light-DM candidates that carry sub-eV kinetic energies, electronic or nuclear excitations employed in operating experiments cannot be induced. However, the deposited energy and momentum are still sufficient to cause a detectable phononic excitation. Here we focus on scattering events, and consider two mechanisms for chiral phonon generation: i) direct excitation, with a corresponding transfer of four-momentum and angular momentum; ii) excitation of an unstable achiral phonon, which has no intrinsic angular momentum, and subsequently decays into two stable chiral phonons with opposite handedness and thus net zero angular momentum.

Many of the possible DM--SMM interactions allow for the transfer of angular momentum to the target material and hence direct generation of chiral phonons. Out of the 18 general operators that can mediate DM interactions, identified within a non-relativistic effective field theory~\cite{Catena:2021qsr}, 14 have non-zero off-diagonal terms of the DM spin component and allow therefore for a spin flip of the incoming particle (as shown in Tab.~\ref{tab:operators}). Similarly, for the case of spin-1 DM candidates all four general operators $\mathcal{O}_{17, 18, 19, 20}$, identified in~\cite{Catena:2019hzw}, can transfer angular momentum to the target.

\begin{table}[t]
    \centering
    \begin{tabular*}{\columnwidth}{@{\extracolsep{\fill}}ll@{}}
    \toprule
       ${\mathcal{O}_4} = \mathbf{S}_{\chi}\cdot \mathbf{S}_t$ & ${\mathcal{O}_{13}} =i \left(\mathbf{S}_{\chi}\cdot  \mathbf{v}^{\perp}_{\rm el}\right)\left(\mathbf{S}_t\cdot \frac{ \mathbf{q}}{m_t}\right)$ \\
        ${\mathcal{O}_5} = i\mathbf{S}_\chi\cdot\left(\frac{ \mathbf{q}}{m_t}\times \mathbf{v}^{\perp}_{\rm el}\right)$ &   ${\mathcal{O}_{14}} = i\left(\mathbf{S}_{\chi}\cdot \frac{ \mathbf{q}}{m_t}\right)\left(\mathbf{S}_t\cdot  \mathbf{v}^{\perp}_{\rm el}\right)$ \\
        ${\mathcal{O}_6} = \left(\mathbf{S}_\chi\cdot\frac{ \mathbf{q}}{m_t}\right) \left(\mathbf{S}_t\cdot\frac{{\bf{q}}}{m_t}\right)$ &   ${\mathcal{O}_{15}} = i\mathcal{O}_{11}\left[ \left(\mathbf{S}_t\times  \mathbf{v}^{\perp}_{\rm el} \right) \cdot \frac{ \mathbf{q}}{m_t}\right] $ \\                                                                             
        ${\mathcal{O}_8} = \mathbf{S}_{\chi}\cdot  \mathbf{v}^{\perp}_{\rm el}$ &  ${\mathcal{O}_{17}} = i\frac{\mathbf{q}}{m_t} \cdot \boldsymbol{\mathcal{S}} \cdot \mathbf{v}^{\perp}_{\rm el} \mathds{1}_{\chi t}$ \\                                                                                                                 
        ${\mathcal{O}_9} = i\mathbf{S}_\chi\cdot\left(\mathbf{S}_t\times\frac{ \mathbf{q}}{m_t}\right)$ &  ${\mathcal{O}_{18}} =i\frac{\mathbf{q}}{m_t} \cdot \boldsymbol{\mathcal{S}} \cdot \mathbf{S}_t$ \\   
        ${\mathcal{O}_{11}} = i\mathbf{S}_\chi\cdot\frac{ \mathbf{q}}{m_t}$ &  ${\mathcal{O}_{19}} =\frac{\mathbf{q}}{m_t} \cdot \boldsymbol{\mathcal{S}} \cdot \frac{\mathbf{q}}{m_t}$ \\
        ${\mathcal{O}_{12}} = \mathbf{S}_{\chi}\cdot \left(\mathbf{S}_t \times \mathbf{v}^{\perp}_{\rm el} \right)$  & ${\mathcal{O}_{20}} = \left(\mathbf{S}_t\times\frac{ \mathbf{q}}{m_t}\right) \cdot \boldsymbol{\mathcal{S}} \cdot \frac{\mathbf{q}}{m_t}$ \\       
    \bottomrule
    \end{tabular*}
    \caption{Interaction operators defining the non-relativistic effective theory of spin-$1/2$ ($\mathcal{O}_{4-15}$) and spin-1 ($\mathcal{O}_{17-20}$) DM-electron interactions~\cite{Fan:2010gt,Fitzpatrick:2012ix,Catena:2019gfa, Catena:2021qsr, Catena:2019hzw}. These can transfer angular momentum to the target and therefore directly cause a chiral phonon excitation via either absorption or spin-flip of the DM particle.~$\mathbf{S}_t$ ($\mathbf{S}_\chi$, $\boldsymbol{\mathcal{S}}$) are the target-particle (DM) spin operators, $\mathbf{q}$ is the transferred momentum, $\mathbf{v}_{\rm el}^\perp$ is the transverse relative velocity, and $m_t$ is the mass of the target-particle (e.g. that of a nucleon or an electron)~\cite{Catena:2019hzw}.}
\label{tab:operators}
\end{table}

An increase of the orbital angular momentum of the atoms creates chiral phonons with a total angular momentum matching that of the excitation (schematically shown in Fig.~\ref{fig-chiral-phonons} \textbf{b}). Even for the spin-diagonal operators or for events where angular momentum is not transferred, in noncentrosymmetric materials chiral phonons can be produced by the creation of an achiral phonon and subsequent decay into two phonons with opposite handedness, conserving the total angular momentum (Fig.~\ref{fig-chiral-phonons} \textbf{c}). Therefore, in a noncentrosymmetric material any DM--SMM interaction that produces phonons can produce chiral phonons.

The maximal energy deposited in a scattering event is the kinetic energy of the incident particle. For the velocities of DM particles expected from the standard halo model (SHM) a chiral phonon detector allows for $\mathcal{O}(\mathrm{keV})$ DM candidate masses to be detected (in contrast with $\mathcal{O}(\mathrm{MeV})$ limits of electronic excitation-based detectors). Since, for the absorption processes, the entire four-momentum of the incident particle is deposited, phononic excitations in this case have the advantage of being able to probe DM candidates with masses down to $\mathcal{O}(\mathrm{meV})$ scales.

Furthermore, absorption of a DM {\it vector} particle always transfers angular momentum to the target. These particles are non-relativistic with a velocity distribution described by  the SHM~\cite{Baxter:2021pqo}, although their interaction and/or production in the Sun can give rise to significant fluxes of relativistic particles here on Earth~\cite{An:2013yua}. Both SHM and boosted relativistic DM absorptions carry four-momenta that are close to the $\Gamma$-point, where they will eventually cause an excitation into a state that is not necessarily an eigenstate, as the eigenstates at $\Gamma$ are achiral. Due to angular momentum conservation, such a state can decay into chiral phonons carrying a net angular momentum given by the spin of the incident particle (Fig.~\ref{fig-chiral-phonons} \textbf{b}).

\section{Detecting chiral phonons} 

The detection of the magnetic moment of a single phonon requires it to be spatially localized for sensing. This is achievable using the phonon Hall effect (Fig.~\ref{fig-chiral-phonons} \textbf{d}), wherein a symmetry-breaking field (\textit{e.g.} a magnetic field, strain, or temperature gradient) causes preferential transport of chiral phonons with opposite handedness in opposite directions, resulting in phonons localized at the surface \cite{zhang2015chiral, park2020phonon}. Engineering the surface crystallographic orientation or geometry can further enhance phonon localization \cite{knipp1989surface}. Since the physical mechanisms of the phonon Hall effect and the electronic Hall effect are related, this approach will also collect any spin-polarized electrons generated by angular momentum transfer from a chiral phonon.  An alternative method of achieving phonon localization could be to use a thin DM detector, restricting them in one dimension \cite{luckyanova2018phonon}.

Once phonons are localized relative to a surface, they can be detected using quantum magnetometers, which are some of the most sensitive experimental probes in existence, able to detect magnetic moments on the order of the Bohr magneton \cite{vasyukov2013scanning, grinolds2013nanoscale}. Coverage of all or part of a surface can be achieved using magnetometer arrays \cite{steinert2010high, drung2007highly}. Unlike a TES, a magnetometer can detect phonons in its proximity and does not require the phonon to cross an interface.

As short-wavelength acoustic phonons in the lowest energy band are stable with respect to decay and host the decay products of higher-energy phonons \cite{maris1993anharmonic}, they are important targets for detection of light dark matter particles. These phonons have submicron wavelengths, and are distinct from the long-wavelength acoustic phonons associated with sound or mechanical oscillation, which could be used to detect wave-like ultralight dark matter \cite{carney2021ultralight}. In a high-purity crystal at millikelvin temperatures, the lifetimes of short-wavelength acoustic phonons are limited by their thermal isolation from the environment. The use of an acoustically bandgapped interface has been demonstrated to achieve excellent thermal isolation, allowing lifetimes $>1$ s \cite{maccabe2020nano}.

We demonstrate below that the maximum value of the phonon angular momentum ($\hbar$) can be approached for the lowest energy phonon band.  Phonon chirality adds conservation of phonon angular momentum as a selection rule for phonon decay processes \cite{pandey2018symmetry}, increasing the phonon lifetimes and ensuring that the anharmonic decay of chiral phonons produces additional chiral phonons, allowing their detection in the acoustic bands.

\section{Materials for DM detection} 

Starting from the requirement that the detector hosts chiral acoustic phonons with large magnetic moments, we identify properties of materials that make them good candidates for chiral phonon-based DM detection. Firstly, the material should lack inversion or time-reversal symmetry \cite{coh2023classification}. We also require materials where the circular motion of atoms has low energy so that the acoustic bands have an associated angular momentum. This requirement suggests the use of crystals whose atoms are topologically underconstrained, allowing, \textit{e.g.}, free rotation in the direction perpendicular to a bond. Such free rotation reduces the energy of modes in which the phonon angular momentum is concentrated on one site, which enhances the phonon magnetic moment, as moments from co-rotating sites of opposite effective charge would point in opposite directions. Consideration of the relative motions of cations and anions can also be relevant in cases where large phonon magnetic moments arise from orbit--lattice coupling, \textit{e.g.} in 4$f$ paramagnets \cite{juraschek2022giant, chaudhary2023giant, luo2023large}; in this case the circular motions of anions relative to a magnetic cation lead to enhanced phonon magnetism and therefore it is desirable for the sublattices to have different circular polarization.

We suggest metal--organic frameworks (MOFs) as candidate materials for the detection of chiral phonons resulting from DM excitations. MOFs are coordination polymers that incorporate both metal ions and organic ligands. The choice of ligand allows control over atomic structure and topology, including the formation of noncentrosymmetric structures which can host chiral phonons \cite{wang2012rational}, leading to a variety of rotational dynamics \cite{gonzalez2019rotational}. MOFs can have high structural anisotropy, including (pseudo) 2-D and 1-D structures, yielding anisotropic phononic properties and thereby high detector sensitivity to the DM wind direction. The chemical flexibility of MOFs allows tailoring of their electronic structures, and thereby incorporation of mechanisms known to cause giant phonon magnetic moments, such as inclusion of rare earth ions \cite{das2017new} and nontrivial electronic topology \cite{deng2021designing, ni2022emergence}. Centimeter-scale single crystals of MOFs have been grown both in both 3D and in 2D, making manufacturing large quantities of these materials technologically feasible \cite{kim2019uniform, kim2020centimeter}.

\section{An example framework material}

\begin{figure*}[ht]
\centering
  \includegraphics[width=1\textwidth,trim=0.25cm 2.75cm 0.5cm 3cm,clip]{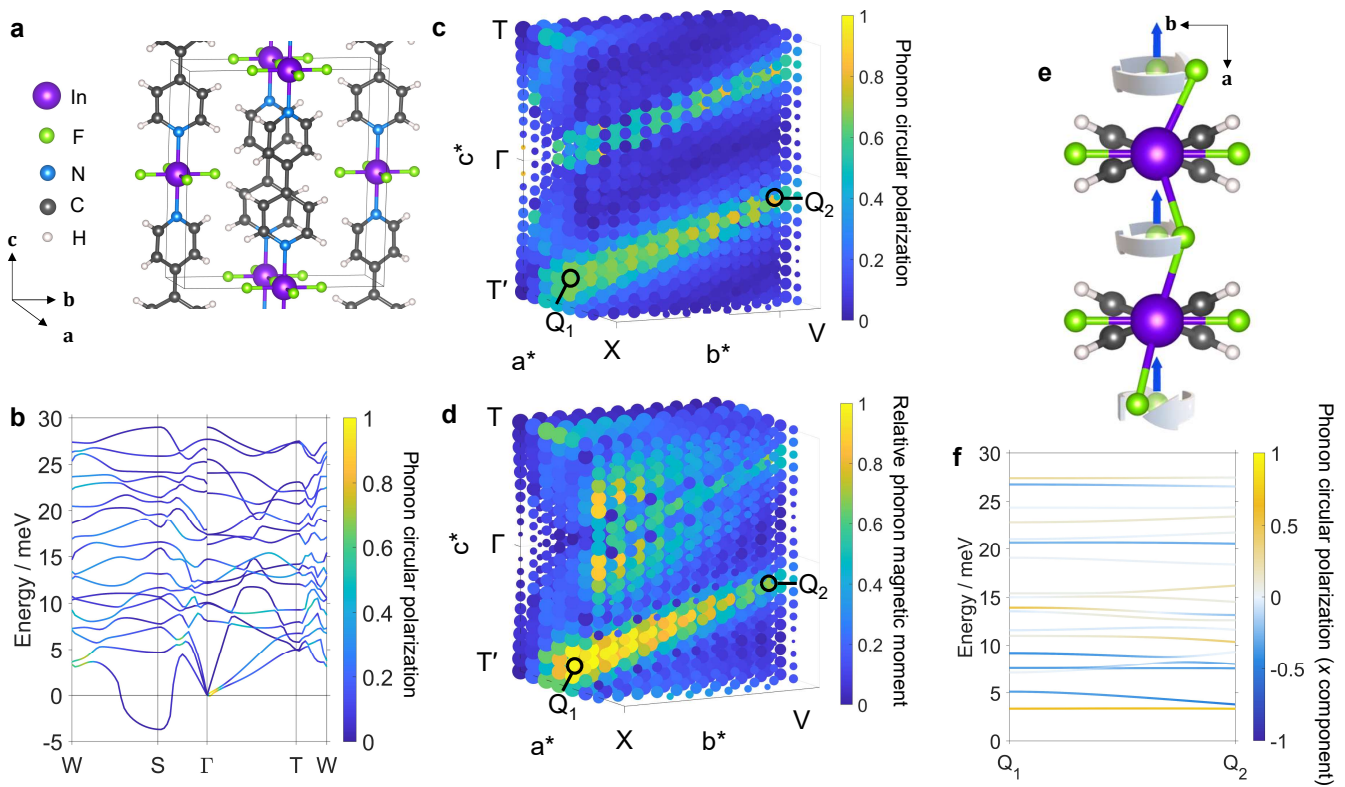}
  \caption{
    \textbf{a} The crystal structure of InF$_3$(bpy), showing In--bpy chains along the $\mathbf{c}$ axis and In--F chains along $\mathbf{a}$ \cite{petrosyants2010organometallic}. The structure is chiral due to the twisting of bipyridine rings relative to each other. \textbf{b} The phonon band structure of InF$_3$(bpy), shown below 30 meV, with bands coloured according to the magnitude of their circular polarization. Special points in and paths through reciprocal space were chosen following Ref. \cite{hinuma2017band}. See footnote \ref{note-imaginary} for discussion of the phonon instability at $\mathbf{S}$ ($0 ~0  ~0.5$). \textbf{c} The lowest energy phonon band in InF$_3$(bpy) shown on a $10 \times 10 \times 20$ grid in reciprocal space, with points coloured by the magnitude of the phonon circular polarization ($||\mathbf{S}|| = ||\mathbf{L}||/\hbar$). Points in reciprocal space are marked $\mathbf{T}$ ($0 ~0 ~0.5$), $\mathbf{T}'$ ($0 ~0 ~-\!0.5$), $\mathbf{V}$ ($0.5  ~-\!0.5  ~-\!0.5$), $\mathbf{X}$ ($0.5 ~0.5  ~-\!0.5$), $\mathbf{\Gamma}$ ($0 ~0 ~0$), $\mathbf{Q}_1 (0.2 ~0 ~-\!0.4)$, and $\mathbf{Q}_2 (0.2 ~0.5 ~-\!0.15)$ (in reduced coordinates). Point sizes are proportional to the phonon energy (modes with imaginary frequencies are not shown). \textbf{d} The lowest energy phonon band shown as in \textbf{c}, coloured by the relative magnitude of their magnetic moment as calculated from the Born effective charge (see text for details), showing the anisotropy of the phonon magnetism. \textbf{e} A schematic view of the motion of the fluorine atoms in the lowest energy acoustic phonon at $\mathbf{Q}_1$, showing how magnetic moments (blue arrows) arise from circular motions (gray arrows) of F anions (green) about their average positions (transparent green spheres), with partial cancellation from smaller circular motions (not shown) of In cations (purple). \textbf{f} The phonon band structure shown on a line connecting $\mathbf{Q}_1$ and $\mathbf{Q}_2$, with bands coloured by the $x$ component of the phonon circular polarization.}
  \label{fig-phonons}
\end{figure*}

As a proof of concept, we screened noncentrosymmetric MOFs using density functional theory (DFT) and density functional perturbation theory (DFPT) to identify materials whose lowest energy phonons are chiral and can carry a magnetic moment, by calculating the phonon eigenvectors, energies, and Born effective charges within the \textsc{Abinit} software package \cite{gonze2020abinit, gonze1997dynamical, bottin2008large, bjorkman2011cif2cell, perdew1996generalized, monkhorst1976special, grimme2010consistent} and using Eq. (\ref{eq-magmo}). This screening identified InF$_3$($4,4'$-bipyridine) (InF$_3$(bpy), space group $\mathrm{I}222$) (Fig.~\ref{fig-phonons} \textbf{a}) \cite{petrosyants2010organometallic} as possessing the desired properties. 

The phononic properties of InF$_3$(bpy) show many of the characteristics needed for DM detection \textit{via} chiral phonons. Its phonon band structure (Fig.~\ref{fig-phonons} \textbf{b}) possesses a high density of low-energy vibrational states.\footnote{\label{note-imaginary}Note that the imaginary phonon frequencies near one face of the Brillouin zone ($\mathbf{c}^\star = \pm0.5$ in reduced coordinates) indicate an instability of the crystal structure with respect to doubling of the $\mathbf{b}$ lattice vector at low temperatures. This instability would reduce the space group symmetry to $C222$, $P222$, $C2$, or $P2$, depending on the order parameter and order parameter direction of the transition, and therefore the structure would remain chiral. It would not be computationally feasible to study this instability due to the large allowed range of $\mathbf{a}^\star$ and $\mathbf{b}^\star$.} Along the high-symmetry directions, the circular polarization is strong in the low-energy ($< 30$ meV) phonons. The phonons are highly anisotropic, developing larger polarizations in regions of reciprocal space away from the high-symmetry directions.  In Fig.~\ref{fig-phonons} \textbf{c} we show the circular polarization of the lowest energy acoustic band on a grid of $\mathbf{q}$-points in reciprocal space. We find a high degree of chirality in specific regions of momentum space. Since DM particles passing through the detector have a preferred direction (opposite to the velocity of the earth in the galactic rest frame), the momentum that they deposit in the material also has a preferred orientation. Consequently, the rate of excitation of chiral phonons will depend on this momentum orientation. Therefore, rotating the detector along and out of this direction will either increase or suppress the observed event rate. Like the phonon angular momentum, the phonon magnetic moment shows a significant degree of anisotropy, allowing high directional sensitivity (Fig.~\ref{fig-phonons} \textbf{d}).

\begin{figure*}[ht]
    \centering
    \includegraphics[width=1\textwidth,trim=0cm 10.5cm 5cm 3cm,clip]{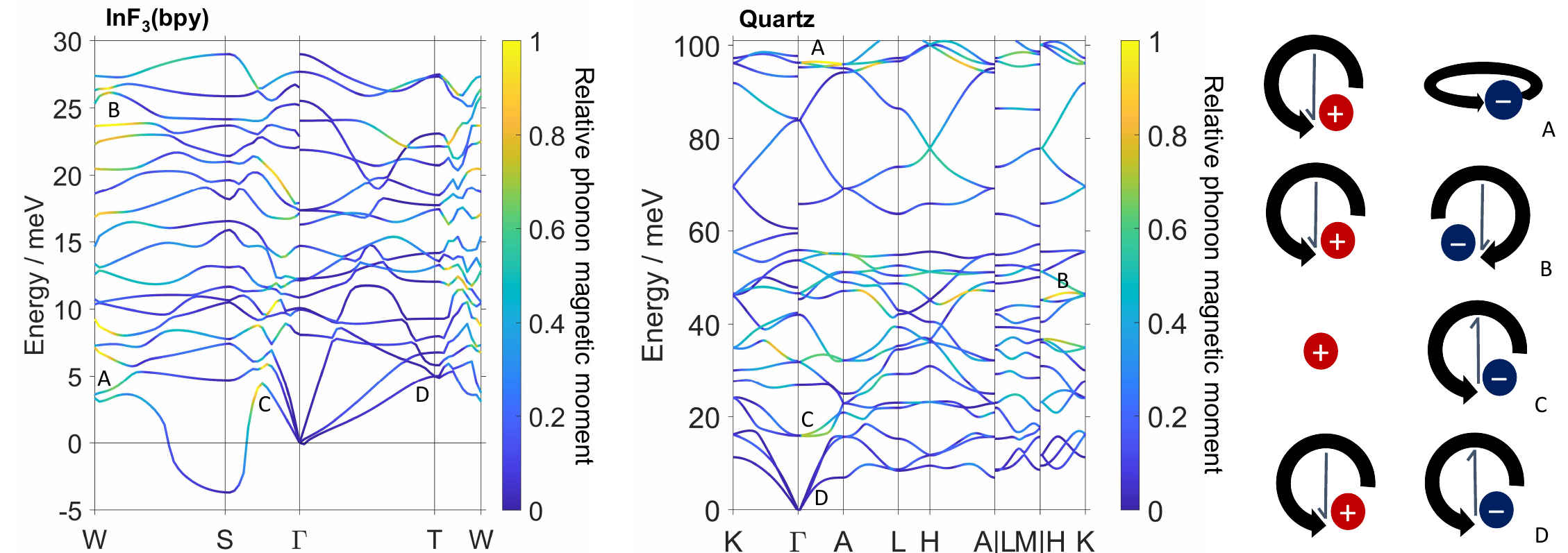}
  \caption{The phonon band structure of InF$_3$(bpy) and $\alpha$-quartz (SiO$_2$), with bands colored according to their relative magnetic moment as calculated from the Born effective charges (see Ref. \cite{quartz} for details). Four types of chiral phonon are marked A--D and shown diagramatically at right: modes where the angular momenta of the cationic and anionic sublattices are orthogonal (A), modes where the momenta are opposed (B), modes where one sublattice is stationary (C), and modes where the momenta are coaligned (D). A, B, and C have relatively large magnetic moments, D does not despite having large angular momentum The MOF structure is more flexible and allows modes with larger magnetic moments in the lowest energy band.
}
  \label{fig-quartz}
\end{figure*}

Notably, the acoustic phonons reach a maximum magnetic moment of 0.02\,$\mu_\mathrm{n}$ within DFT, which is a significant fraction of the maximum magnetic moment of any phonon in any band (0.05\,$\mu_\mathrm{n}$). (Note that, as we calculated the phonon magnetic moments using the currents generated by circular motions of the conventional Born effective charges, they are much smaller than the experimentally measured values ($\sim 1\mu_\mathrm{B}$) discussed above because they do not include contributions from coupling to electronic angular momentum.) The relatively large moments are a result of structural flexibility allowing rotation of fluorine atoms into free space in the crystal structure (Fig.~\ref{fig-phonons} \textbf{e}). The flexibility of the MOF structure is greater than that of simpler framework structures such as quartz, which has relatively smaller magnetic moments in its lowest energy phonon band (Fig.~\ref{fig-quartz}).

Fig.~\ref{fig-phonons} \textbf{f} shows the low-energy phonon band structure along the line of high circular polarization from $\mathbf{Q}_1$ to $\mathbf{Q}_2$. The material possesses a large number of chiral phonons in this energy range which will decay into a chiral acoustic band \cite{pandey2018symmetry}, generating stable  phonons with a net magnetic moment. Fig.~\ref{fig-phonons} \textbf{f} also shows that the lowest energy band has very low dispersion and therefore group velocity in this direction, a common feature of MOFs due to their porosity \cite{kamencek2019understanding}. This indicates spatially localized phonons, which are easier to detect than delocalized phonons.

InF$_3$(bpy) therefore demonstrates that MOFs can host chiral phonons with magnetic moments in their lowest energy phonon bands (with 4\,meV energy threshold).  These moments could be measured at arbitrary points in reciprocal space \textit{via} inelastic neutron scattering with spin-polarized neutrons \cite{squires2012introduction}, allowing screening of materials for their suitability as chiral phonon-based DM detectors.

\section{Reach projections}

\begin{figure*}[ht]
    \centering
    \includegraphics[width=0.48\textwidth]{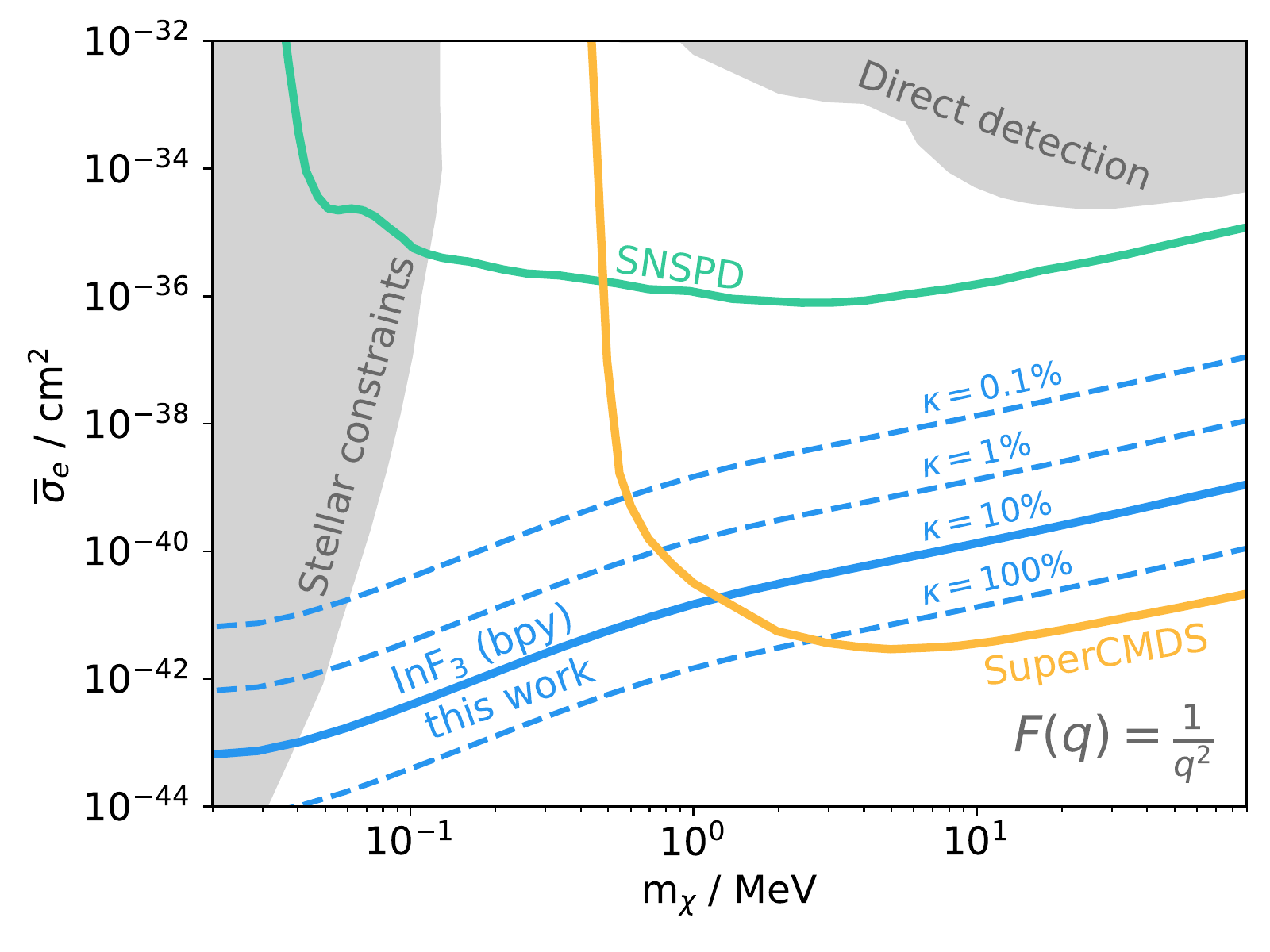}
    \includegraphics[width=0.48\textwidth]{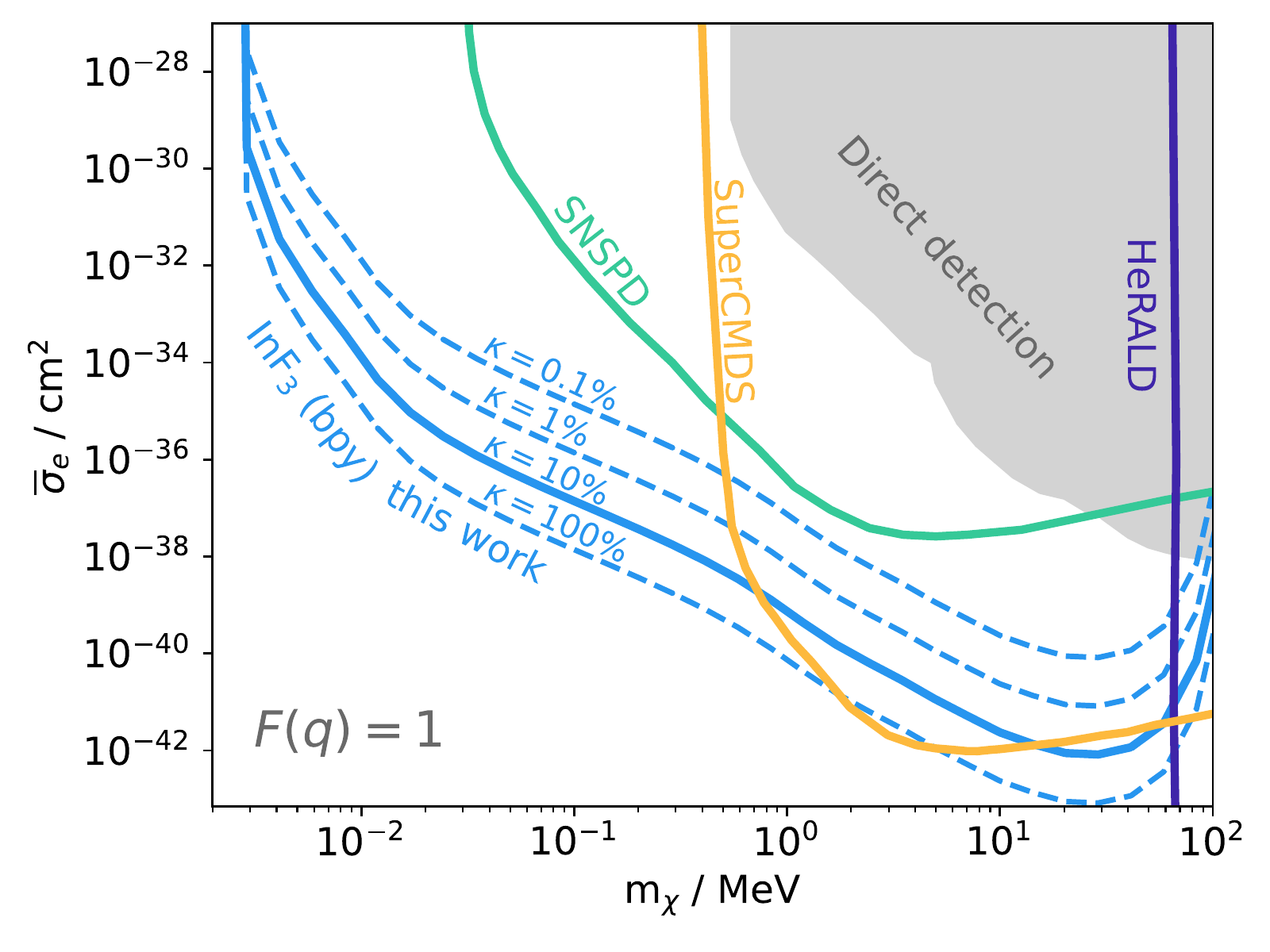}
    \caption{Sensitivity reach for phonon excitations with kg-yr exposures in InF$_3$(bpy) mediated by kinetically mixed light (left) and heavy (right) dark photons calculated in this work compared to other proposed phononic DM detectors. The sensitivity is shown for 0.1\%, 1\%, 10\%, and 100\% probabilities of decay ($\kappa$) into the stable chiral phonon band. Our calculation suggests $\kappa = $15\% and we conservatively highlight the line associated with 10\% probability. The orange line represents the reach of a potential upgrade of the SuperCDMS experiment (silicon) utilizing phonon detection through TES readout~\cite{SuperCDMS:2022kse} of exposure 0.4\,kg-yr.  The green line shows the projected reach of a superconducting NbN nanowire single photon detector (SNSPD) of g-yr exposure as proposed in~\cite{Hochberg:2021yud}. The purple line in the plot shows a helium-based detector (HeRALD, currently in R\&D) that uses phonon-driven atomic ejections from the liquid surface followed by detectable adsorptions for an expected energy threshold of 10\,eV and exposure of a kg-yr~\cite{PhysRevD.100.092007}. Grey-shaded regions are excluded by previous direct detection experiments~\cite{Essig:2015cda, SENSEI:2019ibb, DarkSide:2018ppu, DAMIC:2019dcn} and stellar constraints  taken from~\cite{Chang:2019xva}.}
    \label{fig-phasespace}
\end{figure*}

Finally, we estimate the projected reach of an InF$_3$(bpy) chiral phonon detector with 1\,kg-yr exposure and compare it to other proposed low-energy-threshold detectors (Fig.~\ref{fig-phasespace}). We obtain these projections within the generalized non-relativistic effective operator approach~\cite{Catena:2019gfa} using the dark photon interaction model. In this model, spin-1/2 DM scatters from SMM through an exchange of a kinematically mixed spin-1 boson.

In the non-relativistic limit, this interaction cannot transfer angular momentum to the target as the corresponding spin-matrix element is diagonal in the DM spin space, creating a pair of chiral phonons with opposite angular momenta (Fig.~\ref{fig-chiral-phonons}\textbf{c}).

Excitation rates were obtained starting from the material-specific phonon eigenvectors and eigenenergies obtained from first-principles DFPT calculations and generalized interaction operators from effective field theory. The \mbox{Phono}Dark code was used to combine these and obtain scattering rates within Fermi's golden rule ~\cite{Trickle:2020oki}. By using a truncated boosted Maxwell--Boltzmann distribution to model the DM velocity, we obtained projected detector sensitivity limits. We report these at a 95\% confidence limit assuming zero background (or equivalently three observed events over the considered time period).

We estimate that the fraction of the excited phonons that decay into at least one stable (\textit{i.e.}, having a lifetime sufficiently long to be detectable by a SQUID) chiral product phonon in the lowest energy acoustic band is $\kappa = $\,15\%. This estimate was obtained by taking the expectation value of the circular polarization over all the states in this band ($\kappa = \langle ||\mathbf{S}|| \rangle$), and is conservative in that it does not account for the production of multiple low-energy phonons from higher-energy excitations. We assume that each decay pathway is equally likely, as determination of the probabilities of the decay pathways requires calculation of all three-phonon coupling constants, which is computationally prohibitive for materials with more than a few atoms in the unit cell. We further assume that phonons with relative magnetic moments above 0.4 (\textit{i.e.} green or yellow in Fig.~\ref{fig-phonons} \textbf{d}) are detectable.

When compared to other proposed low-threshold detectors currently under development, such as voltage-biased superconductors~\cite{Hochberg:2021yud}, phononic DM detectors using Si~\cite{SuperCDMS:2022kse} or He~\cite{PhysRevD.100.092007}, or other organic targets~\cite{Blanco:2019lrf, Blanco:2021hlm}, a chiral phonon detector using InF$_3$(bpy) would exhibit greater reach, probing a larger section of unexplored phase space (Fig.~\ref{fig-phasespace}). This reach can be attributed to our proposed method's ability to detect single-phonon excitations and the material's high density of low-energy vibrational states.

We have omitted the projected reach of other proposed (centrosymmetric) single-phonon detector materials (\textit{e.g.} those of Ref. \cite{PhysRevD.101.055004}) as at present the readout of single phonon excitations using a TES is not feasible.  The short decay time (ps) of phonons of energies detectable by a TES ($\sim$ 100 meV) reduces the distance they can travel within a detector to $< 1$\,{\textmu}m, and lower energy phonons do not carry enough energy to create a signal. Furthermore, the transmission probability into the TES is low ($\sim 10^{-3}$)~\cite{SuperCDMS:2022kse} reducing the number of low-energy phonons that can pass the interface and thermalize in the TES. A single excitation of \textit{ca.} 100\,meV is therefore not enough to create a signal and $\mathcal{O} (10)$\,eV of deposited energy in a multitude of phonons is required.

Since current direct detection experiments based on electronic excitations operate with higher detection energy thresholds or $\sim$\,ng-week exposures, constraints from existing measurements have substantially smaller reach. Recently, 2D detector setups with single atomic thickness have been proposed which would operate at the g--kg scale~\cite{Capparelli:2014lua}. If we assume detector mass comparable to those of such detectors, $\mathcal{O}(10)$ orders of improvement with respect to the current limits would be achieved~\cite{Hochberg:2021yud}.

As the DM-induced rate is proportional solely to the total active mass of the detector irrespective of its being one compact object or an array of smaller ones, scalability of the design can be achieved by manufacturing a multitude of detection units of lower mass, should it prove economically more efficient than producing a large bulk sample of the material.

\section{Conclusions}

We have determined that DM--SMM interactions could create chiral phonons either directly, by passing orbital angular momentum to the material through scattering or absorption, or indirectly, by creating phononic excitations that decay into chiral pairs with opposite handedness. Materials with chiral acoustic bands which carry a measurable magnetic moment could then be used to detect DM. Chiral phonons, with their low excitation energy of $\mathcal{O} (\text{meV})$, are ideal probes of light DM candidates down to masses $\mathcal{O} (\text{keV})$ that have not been explored to date. 

By considering the material requirements for a chiral phonon-based DM detector, we have identified MOFs as promising candidates due to their structural flexibility (which allows them to possess low-energy phonons with magnetic moments) and their chemical flexibility (which allows their phonon magnetic moments to be optimized by chemical substitution while maintaining their network topology). Additionally, the anisotropy of MOFs leads to a directional response of the detector that is then able to distinguish common sources of background such as radioactive isotope contamination, cosmic rays and electronic noise. This enables the setup outlined in this work to push the sensitivity of current direct detection experiments by many orders of magnitude exploring both weaker couplings as well as smaller DM candidate masses.

To demonstrate these concepts, we calculated the phonon band structure, circular polarization, and magnetic moments of InF$_3$(bpy), a simple chiral MOF. We found an anisotropic distribution of magnetic-moment carrying chiral phonons in its lowest-energy band, arising from circular librations of In--F bonds, demonstrating that the underconstrained topology typical of MOFs can create the type of phonons needed for DM detection. We then used InF$_3$(bpy) as an example target in order to estimate the experimental reach of a chiral phonon-based detector for the dark photon interaction model, and found that a considerable increase in reach could be achieved if the magnetic moments of individual phonons can be detected.

In future works, it would be valuable to explore the effects of other effective operators on the produced rate, to determine whether DM-interaction differentiation would be possible using this method. From a materials standpoint, the study of the magnetic moments of chiral phonons away from $\Gamma$ is in its infancy, and further experimental and theoretical efforts should focus on understanding the origins of giant phonon magnetic moments, and prediction and measurement of these moments in the acoustic bands of highly anisotropic materials. Additionally, the phonon Hall effect in noncentrosymmetric materials is largely unexplored, and must be studied further, in particular to determine how the symmetry-breaking field and magnetometer readout can be made compatible.

We hope that our study motivates experimental efforts to detect DM using the chiral phonons in MOFs either in thin layers or as a larger bulk sample where the phonon thermal Hall effect is used to localize phonons at the surface.

\section{Methods}

\textbf{Density functional theory:} DFT and DFPT calculations were performed using the \textsc{Abinit} software package (v. 9.6.2)  \cite{gonze2020abinit, gonze1997dynamical, bottin2008large, bjorkman2011cif2cell}. The Perdew--Burke--Ernzerhof exchange--correlation functional \cite{perdew1996generalized} was used with the dispersion correction of Grimme \cite{grimme2010consistent}. A plane-wave basis set (energy cutoff 32 Ha) was used with the default norm-conserving pseudopotentials of the \textsc{Abinit} library. These pseudopotentials contained 36 (In), 2 (F, N, C), and 0 (H) core electrons. Calculations were performed on a $8 \times 8 \times 8$ grid of $\mathbf{k}$-points and a $4 \times 4 \times 4$ Monkhorst--Pack grid \cite{monkhorst1976special} of $\mathbf{q}$-points. The crystal structure was relaxed to an internal pressure of $-0.1$ MPa within the I222 space group prior to DFPT calculations. These calculations were performed for a diamagnetic electronic structure (as In$^{3+}$ is nonmagnetic); we note that coupling between circularly polarized phonons and electronic spins has been successfully modelled (at $\Gamma$) without consideration of modification of the eigenvectors by magnetic effects \cite{juraschek2022giant, chaudhary2023giant, luo2023large}.

\textbf{Excitation rate calculations:}  The PhonoDark software package (v. 1.1.0) ~\cite{Trickle:2020oki} was used to calculate excitation rates; it interfaces Phonopy, an open source package for phonon calculations in Python, with calculations of general effective DM interactions~\cite{Catena:2019gfa}. In order to use results of DFPT calculations from \textsc{Abinit}, the PhonoDark code was modified to accept an input force constant matrix from AbiPy \cite{gonze2020abinit}. An example spin-$1/2$ DM particle interacting through the dark photon model was calculated in the limit of a light mediator (\textit{i.e.}, a long-range interaction with a transferred momentum mediator factor $F(q) = \frac{1}{\mathbf{q}^2}$) and in the limit of a heavy mediator ($F(q) = 1$) following the approach of~\cite{PhysRevD.101.055004}. A fixed time of day of $t=0$ was chosen for an exposure of kg-yr. The exclusion confidence limits were set to 95\% assuming zero background, or equivalently three recorded events. The DM velocity distribution was modeled with a truncated boosted Maxwell--Boltzmann distribution. 

\textbf{Data availability:} All computational data are publicly available from Ref. \cite{sdata}.

\acknowledgements{The authors thank Dieter Kölle, Reinhold Kleiner, and Daniel Bothner for helpful discussions. C.P.R. was supported by ETH Zurich and by the European Union and Horizon 2020 through a Marie Skłodowska-Curie Fellowship, Grant Agreement No. 101030352. R.C. acknowledges support from the Knut and Alice Wallenberg project grant Light Dark Matter (Dnr.~KAW 2019.0080), and from the Swedish Research Council, Dnr. 2018-05029 and Dnr. 2022-04299. N.A.S. and M.M. were supported by ETH Zurich and by the European Research Council (ERC) under the European Union’s Horizon 2020 research and innovation programme project HERO Grant Agreement No. 810451. Computational resources were provided by ETH Zurich and by the Swiss National Supercomputing Center (CSCS) under project IDs s1128 and eth3.}

\bibliography{ref}

\begin{thebibliography}{90}%
\makeatletter
\providecommand \@ifxundefined [1]{%
 \@ifx{#1\undefined}
}%
\providecommand \@ifnum [1]{%
 \ifnum #1\expandafter \@firstoftwo
 \else \expandafter \@secondoftwo
 \fi
}%
\providecommand \@ifx [1]{%
 \ifx #1\expandafter \@firstoftwo
 \else \expandafter \@secondoftwo
 \fi
}%
\providecommand \natexlab [1]{#1}%
\providecommand \enquote  [1]{``#1''}%
\providecommand \bibnamefont  [1]{#1}%
\providecommand \bibfnamefont [1]{#1}%
\providecommand \citenamefont [1]{#1}%
\providecommand \href@noop [0]{\@secondoftwo}%
\providecommand \href [0]{\begingroup \@sanitize@url \@href}%
\providecommand \@href[1]{\@@startlink{#1}\@@href}%
\providecommand \@@href[1]{\endgroup#1\@@endlink}%
\providecommand \@sanitize@url [0]{\catcode `\\12\catcode `\$12\catcode
  `\&12\catcode `\#12\catcode `\^12\catcode `\_12\catcode `\%12\relax}%
\providecommand \@@startlink[1]{}%
\providecommand \@@endlink[0]{}%
\providecommand \url  [0]{\begingroup\@sanitize@url \@url }%
\providecommand \@url [1]{\endgroup\@href {#1}{\urlprefix }}%
\providecommand \urlprefix  [0]{URL }%
\providecommand \Eprint [0]{\href }%
\providecommand \doibase [0]{http://dx.doi.org/}%
\providecommand \selectlanguage [0]{\@gobble}%
\providecommand \bibinfo  [0]{\@secondoftwo}%
\providecommand \bibfield  [0]{\@secondoftwo}%
\providecommand \translation [1]{[#1]}%
\providecommand \BibitemOpen [0]{}%
\providecommand \bibitemStop [0]{}%
\providecommand \bibitemNoStop [0]{.\EOS\space}%
\providecommand \EOS [0]{\spacefactor3000\relax}%
\providecommand \BibitemShut  [1]{\csname bibitem#1\endcsname}%
\let\auto@bib@innerbib\@empty
\bibitem [{\citenamefont {Clowe}\ \emph {et~al.}(2006)\citenamefont {Clowe},
  \citenamefont {Bradac}, \citenamefont {Gonzalez}, \citenamefont {Markevitch},
  \citenamefont {Randall}, \citenamefont {Jones},\ and\ \citenamefont
  {Zaritsky}}]{Clowe:2006eq}%
  \BibitemOpen
  \bibfield  {author} {\bibinfo {author} {\bibfnamefont {Douglas}\ \bibnamefont
  {Clowe}}, \bibinfo {author} {\bibfnamefont {Marusa}\ \bibnamefont {Bradac}},
  \bibinfo {author} {\bibfnamefont {Anthony~H.}\ \bibnamefont {Gonzalez}},
  \bibinfo {author} {\bibfnamefont {Maxim}\ \bibnamefont {Markevitch}},
  \bibinfo {author} {\bibfnamefont {Scott~W.}\ \bibnamefont {Randall}},
  \bibinfo {author} {\bibfnamefont {Christine}\ \bibnamefont {Jones}}, \ and\
  \bibinfo {author} {\bibfnamefont {Dennis}\ \bibnamefont {Zaritsky}},\
  }\bibfield  {title} {\enquote {\bibinfo {title} {{A direct empirical proof of
  the existence of dark matter}},}\ }\href {\doibase 10.1086/508162} {\bibfield
   {journal} {\bibinfo  {journal} {Astrophys. J. Lett.}\ }\textbf {\bibinfo
  {volume} {648}},\ \bibinfo {pages} {L109--L113} (\bibinfo {year} {2006})},\
  \Eprint {http://arxiv.org/abs/astro-ph/0608407} {arXiv:astro-ph/0608407}
  \BibitemShut {NoStop}%
\bibitem [{\citenamefont {Bertone}\ and\ \citenamefont
  {Hooper}(2018)}]{Bertone:2016nfn}%
  \BibitemOpen
  \bibfield  {author} {\bibinfo {author} {\bibfnamefont {Gianfranco}\
  \bibnamefont {Bertone}}\ and\ \bibinfo {author} {\bibfnamefont {Dan}\
  \bibnamefont {Hooper}},\ }\bibfield  {title} {\enquote {\bibinfo {title}
  {{History of dark matter}},}\ }\href {\doibase 10.1103/RevModPhys.90.045002}
  {\bibfield  {journal} {\bibinfo  {journal} {Rev. Mod. Phys.}\ }\textbf
  {\bibinfo {volume} {90}},\ \bibinfo {pages} {045002} (\bibinfo {year}
  {2018})},\ \Eprint {http://arxiv.org/abs/1605.04909} {arXiv:1605.04909
  [astro-ph.CO]} \BibitemShut {NoStop}%
\bibitem [{\citenamefont {Aghanim}\ \emph {et~al.}(2020)\citenamefont {Aghanim}
  \emph {et~al.}}]{Planck:2018vyg}%
  \BibitemOpen
  \bibfield  {author} {\bibinfo {author} {\bibfnamefont {Nabila}\ \bibnamefont
  {Aghanim}} \emph {et~al.} (\bibinfo {collaboration} {Planck}),\ }\bibfield
  {title} {\enquote {\bibinfo {title} {{Planck 2018 results. VI. Cosmological
  parameters}},}\ }\href {\doibase 10.1051/0004-6361/201833910} {\bibfield
  {journal} {\bibinfo  {journal} {Astron. Astrophys.}\ }\textbf {\bibinfo
  {volume} {641}},\ \bibinfo {pages} {A6} (\bibinfo {year} {2020})},\ \bibinfo
  {note} {[Erratum: Astron.Astrophys. 652, C4 (2021)]},\ \Eprint
  {http://arxiv.org/abs/1807.06209} {arXiv:1807.06209 [astro-ph.CO]}
  \BibitemShut {NoStop}%
\bibitem [{\citenamefont {Persic}\ \emph {et~al.}(1996)\citenamefont {Persic},
  \citenamefont {Salucci},\ and\ \citenamefont {Stel}}]{Persic:1995ru}%
  \BibitemOpen
  \bibfield  {author} {\bibinfo {author} {\bibfnamefont {Massimo}\ \bibnamefont
  {Persic}}, \bibinfo {author} {\bibfnamefont {Paolo}\ \bibnamefont {Salucci}},
  \ and\ \bibinfo {author} {\bibfnamefont {Fulvio}\ \bibnamefont {Stel}},\
  }\bibfield  {title} {\enquote {\bibinfo {title} {{The Universal rotation
  curve of spiral galaxies: 1. The Dark matter connection}},}\ }\href {\doibase
  10.1093/mnras/278.1.27} {\bibfield  {journal} {\bibinfo  {journal} {Mon. Not.
  Roy. Astron. Soc.}\ }\textbf {\bibinfo {volume} {281}},\ \bibinfo {pages}
  {27} (\bibinfo {year} {1996})},\ \Eprint
  {http://arxiv.org/abs/astro-ph/9506004} {arXiv:astro-ph/9506004} \BibitemShut
  {NoStop}%
\bibitem [{\citenamefont {Carney}\ \emph {et~al.}(2020)\citenamefont {Carney},
  \citenamefont {Ghosh}, \citenamefont {Krnjaic},\ and\ \citenamefont
  {Taylor}}]{Carney:2019pza}%
  \BibitemOpen
  \bibfield  {author} {\bibinfo {author} {\bibfnamefont {Daniel}\ \bibnamefont
  {Carney}}, \bibinfo {author} {\bibfnamefont {Sohitri}\ \bibnamefont {Ghosh}},
  \bibinfo {author} {\bibfnamefont {Gordan}\ \bibnamefont {Krnjaic}}, \ and\
  \bibinfo {author} {\bibfnamefont {Jacob~M.}\ \bibnamefont {Taylor}},\
  }\bibfield  {title} {\enquote {\bibinfo {title} {{Proposal for gravitational
  direct detection of dark matter}},}\ }\href {\doibase
  10.1103/PhysRevD.102.072003} {\bibfield  {journal} {\bibinfo  {journal}
  {Phys. Rev. D}\ }\textbf {\bibinfo {volume} {102}},\ \bibinfo {pages}
  {072003} (\bibinfo {year} {2020})},\ \Eprint
  {http://arxiv.org/abs/1903.00492} {arXiv:1903.00492 [hep-ph]} \BibitemShut
  {NoStop}%
\bibitem [{\citenamefont {Aalbers}\ \emph {et~al.}(2022)\citenamefont {Aalbers}
  \emph {et~al.}}]{LZ:2022ufs}%
  \BibitemOpen
  \bibfield  {author} {\bibinfo {author} {\bibfnamefont {Jelle}\ \bibnamefont
  {Aalbers}} \emph {et~al.} (\bibinfo {collaboration} {LZ}),\ }\bibfield
  {title} {\enquote {\bibinfo {title} {{First Dark Matter Search Results from
  the LUX-ZEPLIN (LZ) Experiment}},}\ }\href@noop {} {\  (\bibinfo {year}
  {2022})},\ \Eprint {http://arxiv.org/abs/2207.03764} {arXiv:2207.03764
  [hep-ex]} \BibitemShut {NoStop}%
\bibitem [{\citenamefont {Meng}\ \emph {et~al.}(2021)\citenamefont {Meng} \emph
  {et~al.}}]{PandaX-4T:2021bab}%
  \BibitemOpen
  \bibfield  {author} {\bibinfo {author} {\bibfnamefont {Yue}\ \bibnamefont
  {Meng}} \emph {et~al.} (\bibinfo {collaboration} {PandaX-4T}),\ }\bibfield
  {title} {\enquote {\bibinfo {title} {{Dark Matter Search Results from the
  PandaX-4T Commissioning Run}},}\ }\href {\doibase
  10.1103/PhysRevLett.127.261802} {\bibfield  {journal} {\bibinfo  {journal}
  {Phys. Rev. Lett.}\ }\textbf {\bibinfo {volume} {127}},\ \bibinfo {pages}
  {261802} (\bibinfo {year} {2021})},\ \Eprint
  {http://arxiv.org/abs/2107.13438} {arXiv:2107.13438 [hep-ex]} \BibitemShut
  {NoStop}%
\bibitem [{\citenamefont {Aprile}\ \emph {et~al.}(2018)\citenamefont {Aprile}
  \emph {et~al.}}]{XENON:2018voc}%
  \BibitemOpen
  \bibfield  {author} {\bibinfo {author} {\bibfnamefont {Elena}\ \bibnamefont
  {Aprile}} \emph {et~al.} (\bibinfo {collaboration} {XENON}),\ }\bibfield
  {title} {\enquote {\bibinfo {title} {{Dark Matter Search Results from a One
  Ton-Year Exposure of XENON1T}},}\ }\href {\doibase
  10.1103/PhysRevLett.121.111302} {\bibfield  {journal} {\bibinfo  {journal}
  {Phys. Rev. Lett.}\ }\textbf {\bibinfo {volume} {121}},\ \bibinfo {pages}
  {111302} (\bibinfo {year} {2018})},\ \Eprint
  {http://arxiv.org/abs/1805.12562} {arXiv:1805.12562 [astro-ph.CO]}
  \BibitemShut {NoStop}%
\bibitem [{\citenamefont {Battaglieri}\ \emph {et~al.}(2017)\citenamefont
  {Battaglieri} \emph {et~al.}}]{Battaglieri:2017aum}%
  \BibitemOpen
  \bibfield  {author} {\bibinfo {author} {\bibfnamefont {Marco}\ \bibnamefont
  {Battaglieri}} \emph {et~al.},\ }\bibfield  {title} {\enquote {\bibinfo
  {title} {{US Cosmic Visions: New Ideas in Dark Matter 2017; Community
  Report}},}\ }\href@noop {} {\bibfield  {journal} {\bibinfo  {journal}
  {FERMILAB-CONF-17-282-AE-PPD-T}\ } (\bibinfo {year} {2017})},\ \Eprint
  {http://arxiv.org/abs/1707.04591} {arXiv:1707.04591 [hep-ph]} \BibitemShut
  {NoStop}%
\bibitem [{\citenamefont {Hochberg}\ \emph {et~al.}(2021)\citenamefont
  {Hochberg}, \citenamefont {Lehmann}, \citenamefont {Charaev}, \citenamefont
  {Chiles}, \citenamefont {Colangelo}, \citenamefont {Nam},\ and\ \citenamefont
  {Berggren}}]{Hochberg:2021yud}%
  \BibitemOpen
  \bibfield  {author} {\bibinfo {author} {\bibfnamefont {Yonit}\ \bibnamefont
  {Hochberg}}, \bibinfo {author} {\bibfnamefont {Benjamin~V.}\ \bibnamefont
  {Lehmann}}, \bibinfo {author} {\bibfnamefont {Ilya}\ \bibnamefont {Charaev}},
  \bibinfo {author} {\bibfnamefont {Jeff}\ \bibnamefont {Chiles}}, \bibinfo
  {author} {\bibfnamefont {Marco}\ \bibnamefont {Colangelo}}, \bibinfo {author}
  {\bibfnamefont {Sae~Woo}\ \bibnamefont {Nam}}, \ and\ \bibinfo {author}
  {\bibfnamefont {Karl~K.}\ \bibnamefont {Berggren}},\ }\bibfield  {title}
  {\enquote {\bibinfo {title} {{New Constraints on Dark Matter from
  Superconducting Nanowires}},}\ }\href@noop {} {\  (\bibinfo {year} {2021})},\
  \Eprint {http://arxiv.org/abs/2110.01586} {arXiv:2110.01586 [hep-ph]}
  \BibitemShut {NoStop}%
\bibitem [{\citenamefont {Hochberg}\ \emph {et~al.}(2016)\citenamefont
  {Hochberg}, \citenamefont {Lin},\ and\ \citenamefont
  {Zurek}}]{Hochberg:2016ajh}%
  \BibitemOpen
  \bibfield  {author} {\bibinfo {author} {\bibfnamefont {Yonit}\ \bibnamefont
  {Hochberg}}, \bibinfo {author} {\bibfnamefont {Tongyan}\ \bibnamefont {Lin}},
  \ and\ \bibinfo {author} {\bibfnamefont {Kathryn~M.}\ \bibnamefont {Zurek}},\
  }\bibfield  {title} {\enquote {\bibinfo {title} {{Detecting Ultralight
  Bosonic Dark Matter via Absorption in Superconductors}},}\ }\href {\doibase
  10.1103/PhysRevD.94.015019} {\bibfield  {journal} {\bibinfo  {journal} {Phys.
  Rev. D}\ }\textbf {\bibinfo {volume} {94}},\ \bibinfo {pages} {015019}
  (\bibinfo {year} {2016})},\ \Eprint {http://arxiv.org/abs/1604.06800}
  {arXiv:1604.06800 [hep-ph]} \BibitemShut {NoStop}%
\bibitem [{\citenamefont {Hochberg}\ \emph {et~al.}(2017)\citenamefont
  {Hochberg}, \citenamefont {Lin},\ and\ \citenamefont
  {Zurek}}]{Hochberg:2016sqx}%
  \BibitemOpen
  \bibfield  {author} {\bibinfo {author} {\bibfnamefont {Yonit}\ \bibnamefont
  {Hochberg}}, \bibinfo {author} {\bibfnamefont {Tongyan}\ \bibnamefont {Lin}},
  \ and\ \bibinfo {author} {\bibfnamefont {Kathryn~M.}\ \bibnamefont {Zurek}},\
  }\bibfield  {title} {\enquote {\bibinfo {title} {{Absorption of light dark
  matter in semiconductors}},}\ }\href {\doibase 10.1103/PhysRevD.95.023013}
  {\bibfield  {journal} {\bibinfo  {journal} {Phys. Rev. D}\ }\textbf {\bibinfo
  {volume} {95}},\ \bibinfo {pages} {023013} (\bibinfo {year} {2017})},\
  \Eprint {http://arxiv.org/abs/1608.01994} {arXiv:1608.01994 [hep-ph]}
  \BibitemShut {NoStop}%
\bibitem [{\citenamefont {Qian}\ \emph {et~al.}(2021)\citenamefont {Qian},
  \citenamefont {Zhou},\ and\ \citenamefont {Chen}}]{qian2021phonon}%
  \BibitemOpen
  \bibfield  {author} {\bibinfo {author} {\bibfnamefont {Xin}\ \bibnamefont
  {Qian}}, \bibinfo {author} {\bibfnamefont {Jiawei}\ \bibnamefont {Zhou}}, \
  and\ \bibinfo {author} {\bibfnamefont {Gang}\ \bibnamefont {Chen}},\
  }\bibfield  {title} {\enquote {\bibinfo {title} {Phonon-engineered extreme
  thermal conductivity materials},}\ }\href {\doibase s41563-021-00918-3}
  {\bibfield  {journal} {\bibinfo  {journal} {Nat. Mater.}\ }\textbf {\bibinfo
  {volume} {20}},\ \bibinfo {pages} {1188--1202} (\bibinfo {year}
  {2021})}\BibitemShut {NoStop}%
\bibitem [{\citenamefont {Kahn}\ \emph {et~al.}(2021)\citenamefont {Kahn},
  \citenamefont {Krnjaic},\ and\ \citenamefont
  {Mandava}}]{PhysRevLett.127.081804}%
  \BibitemOpen
  \bibfield  {author} {\bibinfo {author} {\bibfnamefont {Yonatan}\ \bibnamefont
  {Kahn}}, \bibinfo {author} {\bibfnamefont {Gordan}\ \bibnamefont {Krnjaic}},
  \ and\ \bibinfo {author} {\bibfnamefont {Bashi}\ \bibnamefont {Mandava}},\
  }\bibfield  {title} {\enquote {\bibinfo {title} {Dark matter detection with
  bound nuclear targets: The {P}oisson phonon tail},}\ }\href {\doibase
  10.1103/PhysRevLett.127.081804} {\bibfield  {journal} {\bibinfo  {journal}
  {Phys. Rev. Lett.}\ }\textbf {\bibinfo {volume} {127}},\ \bibinfo {pages}
  {081804} (\bibinfo {year} {2021})}\BibitemShut {NoStop}%
\bibitem [{\citenamefont {Schutz}\ and\ \citenamefont
  {Zurek}(2016)}]{Schutz:2016tid}%
  \BibitemOpen
  \bibfield  {author} {\bibinfo {author} {\bibfnamefont {Katelin}\ \bibnamefont
  {Schutz}}\ and\ \bibinfo {author} {\bibfnamefont {Kathryn~M.}\ \bibnamefont
  {Zurek}},\ }\bibfield  {title} {\enquote {\bibinfo {title} {{Detectability of
  Light Dark Matter with Superfluid Helium}},}\ }\href {\doibase
  10.1103/PhysRevLett.117.121302} {\bibfield  {journal} {\bibinfo  {journal}
  {Phys. Rev. Lett.}\ }\textbf {\bibinfo {volume} {117}},\ \bibinfo {pages}
  {121302} (\bibinfo {year} {2016})},\ \Eprint
  {http://arxiv.org/abs/1604.08206} {arXiv:1604.08206 [hep-ph]} \BibitemShut
  {NoStop}%
\bibitem [{\citenamefont {Hertel}\ \emph {et~al.}(2019)\citenamefont {Hertel}
  \emph {et~al.}}]{PhysRevD.100.092007}%
  \BibitemOpen
  \bibfield  {author} {\bibinfo {author} {\bibfnamefont {Scott~A.}\
  \bibnamefont {Hertel}} \emph {et~al.},\ }\bibfield  {title} {\enquote
  {\bibinfo {title} {Direct detection of sub-{G}e{V} dark matter using a
  superfluid $^{4}\mathrm{He}$ target},}\ }\href {\doibase
  10.1103/PhysRevD.100.092007} {\bibfield  {journal} {\bibinfo  {journal}
  {Phys. Rev. D}\ }\textbf {\bibinfo {volume} {100}},\ \bibinfo {pages}
  {092007} (\bibinfo {year} {2019})}\BibitemShut {NoStop}%
\bibitem [{\citenamefont {Campbell-Deem}\ \emph {et~al.}(2022)\citenamefont
  {Campbell-Deem}, \citenamefont {Knapen}, \citenamefont {Lin},\ and\
  \citenamefont {Villarama}}]{Campbell-Deem:2022fqm}%
  \BibitemOpen
  \bibfield  {author} {\bibinfo {author} {\bibfnamefont {Brian}\ \bibnamefont
  {Campbell-Deem}}, \bibinfo {author} {\bibfnamefont {Simon}\ \bibnamefont
  {Knapen}}, \bibinfo {author} {\bibfnamefont {Tongyan}\ \bibnamefont {Lin}}, \
  and\ \bibinfo {author} {\bibfnamefont {Ethan}\ \bibnamefont {Villarama}},\
  }\bibfield  {title} {\enquote {\bibinfo {title} {{Dark matter direct
  detection from the single phonon to the nuclear recoil regime}},}\ }\href
  {\doibase 10.1103/PhysRevD.106.036019} {\bibfield  {journal} {\bibinfo
  {journal} {Phys. Rev. D}\ }\textbf {\bibinfo {volume} {106}},\ \bibinfo
  {pages} {036019} (\bibinfo {year} {2022})},\ \Eprint
  {http://arxiv.org/abs/2205.02250} {arXiv:2205.02250 [hep-ph]} \BibitemShut
  {NoStop}%
\bibitem [{\citenamefont {Trickle}\ \emph {et~al.}(2020)\citenamefont
  {Trickle}, \citenamefont {Zhang}, \citenamefont {Zurek}, \citenamefont
  {Inzani},\ and\ \citenamefont {Griffin}}]{Trickle:2019nya}%
  \BibitemOpen
  \bibfield  {author} {\bibinfo {author} {\bibfnamefont {Tanner}\ \bibnamefont
  {Trickle}}, \bibinfo {author} {\bibfnamefont {Zhengkang}\ \bibnamefont
  {Zhang}}, \bibinfo {author} {\bibfnamefont {Kathryn~M.}\ \bibnamefont
  {Zurek}}, \bibinfo {author} {\bibfnamefont {Katherine}\ \bibnamefont
  {Inzani}}, \ and\ \bibinfo {author} {\bibfnamefont {Sin\'ead~M.}\
  \bibnamefont {Griffin}},\ }\bibfield  {title} {\enquote {\bibinfo {title}
  {{Multi-Channel Direct Detection of Light Dark Matter: Theoretical
  Framework}},}\ }\href {\doibase 10.1007/JHEP03(2020)036} {\bibfield
  {journal} {\bibinfo  {journal} {JHEP}\ }\textbf {\bibinfo {volume} {03}},\
  \bibinfo {pages} {036} (\bibinfo {year} {2020})},\ \Eprint
  {http://arxiv.org/abs/1910.08092} {arXiv:1910.08092 [hep-ph]} \BibitemShut
  {NoStop}%
\bibitem [{\citenamefont {Trickle}\ \emph {et~al.}(2022)\citenamefont
  {Trickle}, \citenamefont {Zhang},\ and\ \citenamefont
  {Zurek}}]{Trickle:2020oki}%
  \BibitemOpen
  \bibfield  {author} {\bibinfo {author} {\bibfnamefont {Tanner}\ \bibnamefont
  {Trickle}}, \bibinfo {author} {\bibfnamefont {Zhengkang}\ \bibnamefont
  {Zhang}}, \ and\ \bibinfo {author} {\bibfnamefont {Kathryn~M.}\ \bibnamefont
  {Zurek}},\ }\bibfield  {title} {\enquote {\bibinfo {title} {{Effective field
  theory of dark matter direct detection with collective excitations}},}\
  }\href {\doibase 10.1103/PhysRevD.105.015001} {\bibfield  {journal} {\bibinfo
   {journal} {Phys. Rev. D}\ }\textbf {\bibinfo {volume} {105}},\ \bibinfo
  {pages} {015001} (\bibinfo {year} {2022})},\ \Eprint
  {http://arxiv.org/abs/2009.13534} {arXiv:2009.13534 [hep-ph]} \BibitemShut
  {NoStop}%
\bibitem [{\citenamefont {Griffin}\ \emph {et~al.}(2018)\citenamefont
  {Griffin}, \citenamefont {Knapen}, \citenamefont {Lin},\ and\ \citenamefont
  {Zurek}}]{Griffin:2018bjn}%
  \BibitemOpen
  \bibfield  {author} {\bibinfo {author} {\bibfnamefont {Sinead}\ \bibnamefont
  {Griffin}}, \bibinfo {author} {\bibfnamefont {Simon}\ \bibnamefont {Knapen}},
  \bibinfo {author} {\bibfnamefont {Tongyan}\ \bibnamefont {Lin}}, \ and\
  \bibinfo {author} {\bibfnamefont {Kathryn~M.}\ \bibnamefont {Zurek}},\
  }\bibfield  {title} {\enquote {\bibinfo {title} {{Directional Detection of
  Light Dark Matter with Polar Materials}},}\ }\href {\doibase
  10.1103/PhysRevD.98.115034} {\bibfield  {journal} {\bibinfo  {journal} {Phys.
  Rev. D}\ }\textbf {\bibinfo {volume} {98}},\ \bibinfo {pages} {115034}
  (\bibinfo {year} {2018})},\ \Eprint {http://arxiv.org/abs/1807.10291}
  {arXiv:1807.10291 [hep-ph]} \BibitemShut {NoStop}%
\bibitem [{\citenamefont {Griffin}\ \emph {et~al.}(2020)\citenamefont
  {Griffin}, \citenamefont {Inzani}, \citenamefont {Trickle}, \citenamefont
  {Zhang},\ and\ \citenamefont {Zurek}}]{PhysRevD.101.055004}%
  \BibitemOpen
  \bibfield  {author} {\bibinfo {author} {\bibfnamefont {Sin\'ead~M.}\
  \bibnamefont {Griffin}}, \bibinfo {author} {\bibfnamefont {Katherine}\
  \bibnamefont {Inzani}}, \bibinfo {author} {\bibfnamefont {Tanner}\
  \bibnamefont {Trickle}}, \bibinfo {author} {\bibfnamefont {Zhengkang}\
  \bibnamefont {Zhang}}, \ and\ \bibinfo {author} {\bibfnamefont {Kathryn~M.}\
  \bibnamefont {Zurek}},\ }\bibfield  {title} {\enquote {\bibinfo {title}
  {Multichannel direct detection of light dark matter: Target comparison},}\
  }\href {\doibase 10.1103/PhysRevD.101.055004} {\bibfield  {journal} {\bibinfo
   {journal} {Phys. Rev. D}\ }\textbf {\bibinfo {volume} {101}},\ \bibinfo
  {pages} {055004} (\bibinfo {year} {2020})}\BibitemShut {NoStop}%
\bibitem [{\citenamefont {More}\ and\ \citenamefont {Maris}(1993)}]{NaF_dm}%
  \BibitemOpen
  \bibfield  {author} {\bibinfo {author} {\bibfnamefont {Tamar}\ \bibnamefont
  {More}}\ and\ \bibinfo {author} {\bibfnamefont {Humphrey~J.}\ \bibnamefont
  {Maris}},\ }\bibfield  {title} {\enquote {\bibinfo {title} {Directionality
  from anisotropic phonon production in solid state dark matter detection},}\
  }\href {\doibase 10.1007/BF00693449} {\bibfield  {journal} {\bibinfo
  {journal} {Journal of Low Temperature Physics}\ }\textbf {\bibinfo {volume}
  {93}},\ \bibinfo {pages} {055004} (\bibinfo {year} {1993})}\BibitemShut
  {NoStop}%
\bibitem [{\citenamefont {Coskuner}\ \emph {et~al.}(2022)\citenamefont
  {Coskuner}, \citenamefont {Trickle}, \citenamefont {Zhang},\ and\
  \citenamefont {Zurek}}]{PhysRevD.105.015010}%
  \BibitemOpen
  \bibfield  {author} {\bibinfo {author} {\bibfnamefont {Ahmet}\ \bibnamefont
  {Coskuner}}, \bibinfo {author} {\bibfnamefont {Tanner}\ \bibnamefont
  {Trickle}}, \bibinfo {author} {\bibfnamefont {Zhengkang}\ \bibnamefont
  {Zhang}}, \ and\ \bibinfo {author} {\bibfnamefont {Kathryn~M.}\ \bibnamefont
  {Zurek}},\ }\bibfield  {title} {\enquote {\bibinfo {title} {Directional
  detectability of dark matter with single phonon excitations: Target
  comparison},}\ }\href {\doibase 10.1103/PhysRevD.105.015010} {\bibfield
  {journal} {\bibinfo  {journal} {Phys. Rev. D}\ }\textbf {\bibinfo {volume}
  {105}},\ \bibinfo {pages} {015010} (\bibinfo {year} {2022})}\BibitemShut
  {NoStop}%
\bibitem [{\citenamefont {Knapen}\ \emph {et~al.}(2018)\citenamefont {Knapen},
  \citenamefont {Lin}, \citenamefont {Pyle},\ and\ \citenamefont
  {Zurek}}]{Knapen:2017ekk}%
  \BibitemOpen
  \bibfield  {author} {\bibinfo {author} {\bibfnamefont {Simon}\ \bibnamefont
  {Knapen}}, \bibinfo {author} {\bibfnamefont {Tongyan}\ \bibnamefont {Lin}},
  \bibinfo {author} {\bibfnamefont {Matt}\ \bibnamefont {Pyle}}, \ and\
  \bibinfo {author} {\bibfnamefont {Kathryn~M.}\ \bibnamefont {Zurek}},\
  }\bibfield  {title} {\enquote {\bibinfo {title} {{Detection of Light Dark
  Matter With Optical Phonons in Polar Materials}},}\ }\href {\doibase
  10.1016/j.physletb.2018.08.064} {\bibfield  {journal} {\bibinfo  {journal}
  {Phys. Lett.}\ }\textbf {\bibinfo {volume} {B785}},\ \bibinfo {pages}
  {386--390} (\bibinfo {year} {2018})},\ \Eprint
  {http://arxiv.org/abs/1712.06598} {arXiv:1712.06598 [hep-ph]} \BibitemShut
  {NoStop}%
\bibitem [{\citenamefont {Maris}\ and\ \citenamefont
  {Tamura}(1993)}]{maris1993anharmonic}%
  \BibitemOpen
  \bibfield  {author} {\bibinfo {author} {\bibfnamefont {Humphrey~J.}\
  \bibnamefont {Maris}}\ and\ \bibinfo {author} {\bibfnamefont {Shin-ichiro}\
  \bibnamefont {Tamura}},\ }\bibfield  {title} {\enquote {\bibinfo {title}
  {Anharmonic decay and the propagation of phonons in an isotopically pure
  crystal at low temperatures: {A}pplication to dark-matter detection},}\
  }\href {\doibase 10.1103/PhysRevB.47.727} {\bibfield  {journal} {\bibinfo
  {journal} {Phys. Rev. B}\ }\textbf {\bibinfo {volume} {47}},\ \bibinfo
  {pages} {727} (\bibinfo {year} {1993})}\BibitemShut {NoStop}%
\bibitem [{\citenamefont {Abdelhameed}\ \emph {et~al.}(2019)\citenamefont
  {Abdelhameed} \emph {et~al.}}]{CRESST:2019jnq}%
  \BibitemOpen
  \bibfield  {author} {\bibinfo {author} {\bibfnamefont {Ahmed~H.}\
  \bibnamefont {Abdelhameed}} \emph {et~al.} (\bibinfo {collaboration}
  {CRESST}),\ }\bibfield  {title} {\enquote {\bibinfo {title} {{First results
  from the CRESST-III low-mass dark matter program}},}\ }\href {\doibase
  10.1103/PhysRevD.100.102002} {\bibfield  {journal} {\bibinfo  {journal}
  {Phys. Rev. D}\ }\textbf {\bibinfo {volume} {100}},\ \bibinfo {pages}
  {102002} (\bibinfo {year} {2019})},\ \Eprint
  {http://arxiv.org/abs/1904.00498} {arXiv:1904.00498 [astro-ph.CO]}
  \BibitemShut {NoStop}%
\bibitem [{\citenamefont {Albakry}\ \emph {et~al.}(2022)\citenamefont {Albakry}
  \emph {et~al.}}]{SuperCDMS:2022kse}%
  \BibitemOpen
  \bibfield  {author} {\bibinfo {author} {\bibfnamefont {Musaab~F.}\
  \bibnamefont {Albakry}} \emph {et~al.} (\bibinfo {collaboration}
  {SuperCDMS}),\ }\bibfield  {title} {\enquote {\bibinfo {title} {{A Strategy
  for Low-Mass Dark Matter Searches with Cryogenic Detectors in the SuperCDMS
  SNOLAB Facility}},}\ }in\ \href@noop {} {\emph {\bibinfo {booktitle} {{2022
  Snowmass Summer Study}}}}\ (\bibinfo {year} {2022})\ \Eprint
  {http://arxiv.org/abs/2203.08463} {arXiv:2203.08463 [physics.ins-det]}
  \BibitemShut {NoStop}%
\bibitem [{\citenamefont {Zhang}\ and\ \citenamefont
  {Niu}(2015)}]{zhang2015chiral}%
  \BibitemOpen
  \bibfield  {author} {\bibinfo {author} {\bibfnamefont {Lifa}\ \bibnamefont
  {Zhang}}\ and\ \bibinfo {author} {\bibfnamefont {Qian}\ \bibnamefont {Niu}},\
  }\bibfield  {title} {\enquote {\bibinfo {title} {Chiral phonons at
  high-symmetry points in monolayer hexagonal lattices},}\ }\href {\doibase
  10.1103/PhysRevLett.115.115502} {\bibfield  {journal} {\bibinfo  {journal}
  {Phys. Rev. Lett.}\ }\textbf {\bibinfo {volume} {115}},\ \bibinfo {pages}
  {115502} (\bibinfo {year} {2015})}\BibitemShut {NoStop}%
\bibitem [{\citenamefont {Schaack}(1976)}]{schaack1976observation}%
  \BibitemOpen
  \bibfield  {author} {\bibinfo {author} {\bibfnamefont {Gerhard}\ \bibnamefont
  {Schaack}},\ }\bibfield  {title} {\enquote {\bibinfo {title} {Observation of
  circularly polarized phonon states in an external magnetic field},}\ }\href
  {\doibase 10.1088/0022-3719/9/11/009} {\bibfield  {journal} {\bibinfo
  {journal} {J. Phys. C: Solid State Phys.}\ }\textbf {\bibinfo {volume} {9}},\
  \bibinfo {pages} {L297} (\bibinfo {year} {1976})}\BibitemShut {NoStop}%
\bibitem [{\citenamefont {Baydin}\ \emph {et~al.}(2022)\citenamefont {Baydin},
  \citenamefont {Hernandez}, \citenamefont {Rodriguez-Vega}, \citenamefont
  {Okazaki}, \citenamefont {Tay}, \citenamefont {Noe}, \citenamefont
  {Katayama}, \citenamefont {Takeda}, \citenamefont {Nojiri}, \citenamefont
  {Rappl}, \citenamefont {Abramof}, \citenamefont {Fiete},\ and\ \citenamefont
  {Kono}}]{BaydinPbTe}%
  \BibitemOpen
  \bibfield  {author} {\bibinfo {author} {\bibfnamefont {Andrey}\ \bibnamefont
  {Baydin}}, \bibinfo {author} {\bibfnamefont {Felix G.~G.}\ \bibnamefont
  {Hernandez}}, \bibinfo {author} {\bibfnamefont {Martin}\ \bibnamefont
  {Rodriguez-Vega}}, \bibinfo {author} {\bibfnamefont {Anderson~K.}\
  \bibnamefont {Okazaki}}, \bibinfo {author} {\bibfnamefont {Fuyang}\
  \bibnamefont {Tay}}, \bibinfo {author} {\bibfnamefont {G.~Timothy}\
  \bibnamefont {Noe}}, \bibinfo {author} {\bibfnamefont {Ikufumi}\ \bibnamefont
  {Katayama}}, \bibinfo {author} {\bibfnamefont {Jun}\ \bibnamefont {Takeda}},
  \bibinfo {author} {\bibfnamefont {Hiroyuki}\ \bibnamefont {Nojiri}}, \bibinfo
  {author} {\bibfnamefont {Paulo H.~O.}\ \bibnamefont {Rappl}}, \bibinfo
  {author} {\bibfnamefont {Eduardo}\ \bibnamefont {Abramof}}, \bibinfo {author}
  {\bibfnamefont {Gregory~A.}\ \bibnamefont {Fiete}}, \ and\ \bibinfo {author}
  {\bibfnamefont {Junichiro}\ \bibnamefont {Kono}},\ }\bibfield  {title}
  {\enquote {\bibinfo {title} {Magnetic control of soft chiral phonons in
  {P}b{T}e},}\ }\href {\doibase 10.1103/PhysRevLett.128.075901} {\bibfield
  {journal} {\bibinfo  {journal} {Phys. Rev. Lett.}\ }\textbf {\bibinfo
  {volume} {128}},\ \bibinfo {pages} {075901} (\bibinfo {year}
  {2022})}\BibitemShut {NoStop}%
\bibitem [{\citenamefont {Cheng}\ \emph {et~al.}(2020)\citenamefont {Cheng},
  \citenamefont {Schumann}, \citenamefont {Wang}, \citenamefont {Zhang},
  \citenamefont {Barbalas}, \citenamefont {Stemmer},\ and\ \citenamefont
  {Armitage}}]{cheng2020large}%
  \BibitemOpen
  \bibfield  {author} {\bibinfo {author} {\bibfnamefont {Bing}\ \bibnamefont
  {Cheng}}, \bibinfo {author} {\bibfnamefont {T.}~\bibnamefont {Schumann}},
  \bibinfo {author} {\bibfnamefont {Youcheng}\ \bibnamefont {Wang}}, \bibinfo
  {author} {\bibfnamefont {X.}~\bibnamefont {Zhang}}, \bibinfo {author}
  {\bibfnamefont {D.}~\bibnamefont {Barbalas}}, \bibinfo {author}
  {\bibfnamefont {S.}~\bibnamefont {Stemmer}}, \ and\ \bibinfo {author}
  {\bibfnamefont {N.~P.}\ \bibnamefont {Armitage}},\ }\bibfield  {title}
  {\enquote {\bibinfo {title} {A large effective phonon magnetic moment in a
  {D}irac semimetal},}\ }\href {\doibase 10.1021/acs.nanolett.0c01983}
  {\bibfield  {journal} {\bibinfo  {journal} {Nano Lett.}\ }\textbf {\bibinfo
  {volume} {20}},\ \bibinfo {pages} {5991--5996} (\bibinfo {year}
  {2020})}\BibitemShut {NoStop}%
\bibitem [{\citenamefont {Hernandez}\ \emph {et~al.}(2022)\citenamefont
  {Hernandez}, \citenamefont {Baydin}, \citenamefont {Chaudhary}, \citenamefont
  {Tay}, \citenamefont {Katayama}, \citenamefont {Takeda}, \citenamefont
  {Nojiri}, \citenamefont {Okazaki}, \citenamefont {Rappl}, \citenamefont
  {Abramof} \emph {et~al.}}]{hernandez2022chiral}%
  \BibitemOpen
  \bibfield  {author} {\bibinfo {author} {\bibfnamefont {Felix G.~G.}\
  \bibnamefont {Hernandez}}, \bibinfo {author} {\bibfnamefont {Andrey}\
  \bibnamefont {Baydin}}, \bibinfo {author} {\bibfnamefont {Swati}\
  \bibnamefont {Chaudhary}}, \bibinfo {author} {\bibfnamefont {Fuyang}\
  \bibnamefont {Tay}}, \bibinfo {author} {\bibfnamefont {Ikufumi}\ \bibnamefont
  {Katayama}}, \bibinfo {author} {\bibfnamefont {Jun}\ \bibnamefont {Takeda}},
  \bibinfo {author} {\bibfnamefont {Hiroyuki}\ \bibnamefont {Nojiri}}, \bibinfo
  {author} {\bibfnamefont {Anderson~K.}\ \bibnamefont {Okazaki}}, \bibinfo
  {author} {\bibfnamefont {Paulo H.~O.}\ \bibnamefont {Rappl}}, \bibinfo
  {author} {\bibfnamefont {Eduardo}\ \bibnamefont {Abramof}},  \emph {et~al.},\
  }\bibfield  {title} {\enquote {\bibinfo {title} {Chiral phonons with giant
  magnetic moments in a topological crystalline insulator},}\ }\href@noop {} {\
   (\bibinfo {year} {2022})},\ \Eprint {http://arxiv.org/abs/2208.12235}
  {arXiv:2208.12235 [cond-mat.mes-hall]} \BibitemShut {NoStop}%
\bibitem [{\citenamefont {Juraschek}\ \emph {et~al.}(2022)\citenamefont
  {Juraschek}, \citenamefont {Neuman},\ and\ \citenamefont
  {Narang}}]{juraschek2022giant}%
  \BibitemOpen
  \bibfield  {author} {\bibinfo {author} {\bibfnamefont {Dominik~M.}\
  \bibnamefont {Juraschek}}, \bibinfo {author} {\bibfnamefont
  {Tom{\'a}{\v{s}}}\ \bibnamefont {Neuman}}, \ and\ \bibinfo {author}
  {\bibfnamefont {Prineha}\ \bibnamefont {Narang}},\ }\bibfield  {title}
  {\enquote {\bibinfo {title} {Giant effective magnetic fields from optically
  driven chiral phonons in 4{\textit{f}} paramagnets},}\ }\href {\doibase
  10.1103/PhysRevResearch.4.013129} {\bibfield  {journal} {\bibinfo  {journal}
  {Phys. Rev. Res.}\ }\textbf {\bibinfo {volume} {4}},\ \bibinfo {pages}
  {013129} (\bibinfo {year} {2022})}\BibitemShut {NoStop}%
\bibitem [{\citenamefont {Basini}\ \emph {et~al.}(2022)\citenamefont {Basini},
  \citenamefont {Pancaldi}, \citenamefont {Wehinger}, \citenamefont {Udina},
  \citenamefont {Tadano}, \citenamefont {Hoffmann}, \citenamefont {Balatsky},\
  and\ \citenamefont {Bonetti}}]{basini2022terahertz}%
  \BibitemOpen
  \bibfield  {author} {\bibinfo {author} {\bibfnamefont {Martina}\ \bibnamefont
  {Basini}}, \bibinfo {author} {\bibfnamefont {M.}~\bibnamefont {Pancaldi}},
  \bibinfo {author} {\bibfnamefont {B.}~\bibnamefont {Wehinger}}, \bibinfo
  {author} {\bibfnamefont {M.}~\bibnamefont {Udina}}, \bibinfo {author}
  {\bibfnamefont {T.}~\bibnamefont {Tadano}}, \bibinfo {author} {\bibfnamefont
  {M.C.}\ \bibnamefont {Hoffmann}}, \bibinfo {author} {\bibfnamefont {A.V.}\
  \bibnamefont {Balatsky}}, \ and\ \bibinfo {author} {\bibfnamefont
  {S.}~\bibnamefont {Bonetti}},\ }\bibfield  {title} {\enquote {\bibinfo
  {title} {Terahertz electric-field driven dynamical multiferroicity in
  {SrTiO$_3$}},}\ }\href@noop {} {\  (\bibinfo {year} {2022})},\ \Eprint
  {http://arxiv.org/abs/2210.01690} {arXiv:2210.01690 [cond-mat.str-el]}
  \BibitemShut {NoStop}%
\bibitem [{\citenamefont {Luo}\ \emph {et~al.}(2023)\citenamefont {Luo},
  \citenamefont {Lin}, \citenamefont {Zhang}, \citenamefont {Chen},
  \citenamefont {Blackert}, \citenamefont {Xu}, \citenamefont {Yakobson},\ and\
  \citenamefont {Zhu}}]{luo2023large}%
  \BibitemOpen
  \bibfield  {author} {\bibinfo {author} {\bibfnamefont {Jiaming}\ \bibnamefont
  {Luo}}, \bibinfo {author} {\bibfnamefont {Tong}\ \bibnamefont {Lin}},
  \bibinfo {author} {\bibfnamefont {Junjie}\ \bibnamefont {Zhang}}, \bibinfo
  {author} {\bibfnamefont {Xiaotong}\ \bibnamefont {Chen}}, \bibinfo {author}
  {\bibfnamefont {Elizabeth~R.}\ \bibnamefont {Blackert}}, \bibinfo {author}
  {\bibfnamefont {Rui}\ \bibnamefont {Xu}}, \bibinfo {author} {\bibfnamefont
  {Boris~I.}\ \bibnamefont {Yakobson}}, \ and\ \bibinfo {author} {\bibfnamefont
  {Hanyu}\ \bibnamefont {Zhu}},\ }\bibfield  {title} {\enquote {\bibinfo
  {title} {Large effective magnetic fields from chiral phonons in rare-earth
  halides},}\ }\href {\doibase 10.1126/science.adi9601} {\bibfield  {journal}
  {\bibinfo  {journal} {Science}\ }\textbf {\bibinfo {volume} {382}},\ \bibinfo
  {pages} {698--–702} (\bibinfo {year} {2023})},\ \Eprint
  {http://arxiv.org/abs/2306.03852} {arXiv:2306.03852 [cond-mat.mtrl-sci]}
  \BibitemShut {NoStop}%
\bibitem [{\citenamefont {Park}\ and\ \citenamefont
  {Yang}(2020)}]{park2020phonon}%
  \BibitemOpen
  \bibfield  {author} {\bibinfo {author} {\bibfnamefont {Sungjoon}\
  \bibnamefont {Park}}\ and\ \bibinfo {author} {\bibfnamefont {Bohm-Jung}\
  \bibnamefont {Yang}},\ }\bibfield  {title} {\enquote {\bibinfo {title}
  {Phonon angular momentum {H}all effect},}\ }\href {\doibase
  10.1021/acs.nanolett.0c03220} {\bibfield  {journal} {\bibinfo  {journal}
  {Nano Lett.}\ }\textbf {\bibinfo {volume} {20}},\ \bibinfo {pages}
  {7694--7699} (\bibinfo {year} {2020})}\BibitemShut {NoStop}%
\bibitem [{\citenamefont {Coh}(2023)}]{coh2023classification}%
  \BibitemOpen
  \bibfield  {author} {\bibinfo {author} {\bibfnamefont {Sinisa}\ \bibnamefont
  {Coh}},\ }\bibfield  {title} {\enquote {\bibinfo {title} {Classification of
  materials with phonon angular momentum and microscopic origin of angular
  momentum},}\ }\href {\doibase 10.1103/PhysRevB.108.134307} {\bibfield
  {journal} {\bibinfo  {journal} {Physical Review B}\ }\textbf {\bibinfo
  {volume} {108}},\ \bibinfo {pages} {134307} (\bibinfo {year} {2023})},\
  \Eprint {http://arxiv.org/abs/1911.05064} {arXiv:1911.05064
  [cond-mat.mtrl-sci]} \BibitemShut {NoStop}%
\bibitem [{\citenamefont {Juraschek}\ \emph {et~al.}(2017)\citenamefont
  {Juraschek}, \citenamefont {Fechner}, \citenamefont {Balatsky},\ and\
  \citenamefont {Spaldin}}]{juraschek2017dynamical}%
  \BibitemOpen
  \bibfield  {author} {\bibinfo {author} {\bibfnamefont {Dominik~M.}\
  \bibnamefont {Juraschek}}, \bibinfo {author} {\bibfnamefont {Michael}\
  \bibnamefont {Fechner}}, \bibinfo {author} {\bibfnamefont {Alexander~V.}\
  \bibnamefont {Balatsky}}, \ and\ \bibinfo {author} {\bibfnamefont
  {Nicola~A.}\ \bibnamefont {Spaldin}},\ }\bibfield  {title} {\enquote
  {\bibinfo {title} {Dynamical multiferroicity},}\ }\href {\doibase
  10.1103/PhysRevMaterials.1.014401} {\bibfield  {journal} {\bibinfo  {journal}
  {Phys. Rev. Mater.}\ }\textbf {\bibinfo {volume} {1}},\ \bibinfo {pages}
  {014401} (\bibinfo {year} {2017})}\BibitemShut {NoStop}%
\bibitem [{\citenamefont {Juraschek}\ and\ \citenamefont
  {Spaldin}(2019)}]{juraschek2019orbital}%
  \BibitemOpen
  \bibfield  {author} {\bibinfo {author} {\bibfnamefont {Dominik~M.}\
  \bibnamefont {Juraschek}}\ and\ \bibinfo {author} {\bibfnamefont {Nicola~A.}\
  \bibnamefont {Spaldin}},\ }\bibfield  {title} {\enquote {\bibinfo {title}
  {Orbital magnetic moments of phonons},}\ }\href {\doibase
  10.1103/PhysRevMaterials.3.064405} {\bibfield  {journal} {\bibinfo  {journal}
  {Phys. Rev. Mater.}\ }\textbf {\bibinfo {volume} {3}},\ \bibinfo {pages}
  {064405} (\bibinfo {year} {2019})}\BibitemShut {NoStop}%
\bibitem [{\citenamefont {Zhang}\ and\ \citenamefont
  {Niu}(2014)}]{zhang2014angular}%
  \BibitemOpen
  \bibfield  {author} {\bibinfo {author} {\bibfnamefont {Lifa}\ \bibnamefont
  {Zhang}}\ and\ \bibinfo {author} {\bibfnamefont {Qian}\ \bibnamefont {Niu}},\
  }\bibfield  {title} {\enquote {\bibinfo {title} {Angular momentum of phonons
  and the {E}instein--de {H}aas effect},}\ }\href {\doibase
  10.1103/PhysRevLett.112.085503} {\bibfield  {journal} {\bibinfo  {journal}
  {Phys. Rev. Lett.}\ }\textbf {\bibinfo {volume} {112}},\ \bibinfo {pages}
  {085503} (\bibinfo {year} {2014})}\BibitemShut {NoStop}%
\bibitem [{\citenamefont {Riseborough}(2010)}]{riseborough2010quantum}%
  \BibitemOpen
  \bibfield  {author} {\bibinfo {author} {\bibfnamefont {Peter~S.}\
  \bibnamefont {Riseborough}},\ }\bibfield  {title} {\enquote {\bibinfo {title}
  {Quantum fluctuations in insulating ferroelectrics},}\ }\href {\doibase
  10.1016/j.chemphys.2010.02.025} {\bibfield  {journal} {\bibinfo  {journal}
  {Chem. Phys.}\ }\textbf {\bibinfo {volume} {375}},\ \bibinfo {pages}
  {184--186} (\bibinfo {year} {2010})}\BibitemShut {NoStop}%
\bibitem [{\citenamefont {Boyd}\ \emph {et~al.}(2022)\citenamefont {Boyd},
  \citenamefont {Hochberg}, \citenamefont {Kahn}, \citenamefont {Kramer},
  \citenamefont {Kurinsky}, \citenamefont {Lehmann},\ and\ \citenamefont
  {Yu}}]{Boyd:2022tcn}%
  \BibitemOpen
  \bibfield  {author} {\bibinfo {author} {\bibfnamefont {Christian}\
  \bibnamefont {Boyd}}, \bibinfo {author} {\bibfnamefont {Yonit}\ \bibnamefont
  {Hochberg}}, \bibinfo {author} {\bibfnamefont {Yonatan}\ \bibnamefont
  {Kahn}}, \bibinfo {author} {\bibfnamefont {Eric~David}\ \bibnamefont
  {Kramer}}, \bibinfo {author} {\bibfnamefont {Noah}\ \bibnamefont {Kurinsky}},
  \bibinfo {author} {\bibfnamefont {Benjamin~V.}\ \bibnamefont {Lehmann}}, \
  and\ \bibinfo {author} {\bibfnamefont {To~Chin}\ \bibnamefont {Yu}},\
  }\bibfield  {title} {\enquote {\bibinfo {title} {{Directional detection of
  dark matter with anisotropic response functions}},}\ }\href@noop {} {\
  (\bibinfo {year} {2022})},\ \Eprint {http://arxiv.org/abs/2212.04505}
  {arXiv:2212.04505 [hep-ph]} \BibitemShut {NoStop}%
\bibitem [{\citenamefont {Geilhufe}\ and\ \citenamefont
  {Hergert}(2023)}]{geilhufe2023electron}%
  \BibitemOpen
  \bibfield  {author} {\bibinfo {author} {\bibfnamefont {R.~Matthias}\
  \bibnamefont {Geilhufe}}\ and\ \bibinfo {author} {\bibfnamefont {Wolfram}\
  \bibnamefont {Hergert}},\ }\bibfield  {title} {\enquote {\bibinfo {title}
  {Electron magnetic moment of transient chiral phonons in {K}{T}a{O}$_3$},}\
  }\href {\doibase 10.1103/PhysRevB.107.L020406} {\bibfield  {journal}
  {\bibinfo  {journal} {Phys. Rev. B}\ }\textbf {\bibinfo {volume} {107}},\
  \bibinfo {pages} {L020406} (\bibinfo {year} {2023})}\BibitemShut {NoStop}%
\bibitem [{\citenamefont {Ren}\ \emph {et~al.}(2021)\citenamefont {Ren},
  \citenamefont {Xiao}, \citenamefont {Saparov},\ and\ \citenamefont
  {Niu}}]{ren2021phonon}%
  \BibitemOpen
  \bibfield  {author} {\bibinfo {author} {\bibfnamefont {Yafei}\ \bibnamefont
  {Ren}}, \bibinfo {author} {\bibfnamefont {Cong}\ \bibnamefont {Xiao}},
  \bibinfo {author} {\bibfnamefont {Daniyar}\ \bibnamefont {Saparov}}, \ and\
  \bibinfo {author} {\bibfnamefont {Qian}\ \bibnamefont {Niu}},\ }\bibfield
  {title} {\enquote {\bibinfo {title} {Phonon magnetic moment from electronic
  topological magnetization},}\ }\href {\doibase
  10.1103/PhysRevLett.127.186403} {\bibfield  {journal} {\bibinfo  {journal}
  {Phys. Rev. Lett.}\ }\textbf {\bibinfo {volume} {127}},\ \bibinfo {pages}
  {186403} (\bibinfo {year} {2021})}\BibitemShut {NoStop}%
\bibitem [{\citenamefont {Ueda}\ \emph {et~al.}(2023)\citenamefont {Ueda},
  \citenamefont {García-Fernández}, \citenamefont {Agrestini}, \citenamefont
  {Romao}, \citenamefont {van~den Brink}, \citenamefont {Spaldin},
  \citenamefont {Zhou},\ and\ \citenamefont {Staub}}]{quartz}%
  \BibitemOpen
  \bibfield  {author} {\bibinfo {author} {\bibfnamefont {Hiroki}\ \bibnamefont
  {Ueda}}, \bibinfo {author} {\bibfnamefont {Mirian}\ \bibnamefont
  {García-Fernández}}, \bibinfo {author} {\bibfnamefont {Stefano}\
  \bibnamefont {Agrestini}}, \bibinfo {author} {\bibfnamefont {Carl~P.}\
  \bibnamefont {Romao}}, \bibinfo {author} {\bibfnamefont {Jeroen}\
  \bibnamefont {van~den Brink}}, \bibinfo {author} {\bibfnamefont {Nicola~A.}\
  \bibnamefont {Spaldin}}, \bibinfo {author} {\bibfnamefont {Ke-Jin}\
  \bibnamefont {Zhou}}, \ and\ \bibinfo {author} {\bibfnamefont {Urs}\
  \bibnamefont {Staub}},\ }\bibfield  {title} {\enquote {\bibinfo {title}
  {Chiral phonons probed by {X} rays},}\ }\href {\doibase s41586-023-06016-5}
  {\bibfield  {journal} {\bibinfo  {journal} {Nature}\ }\textbf {\bibinfo
  {volume} {618}},\ \bibinfo {pages} {946--950} (\bibinfo {year} {2023})},\
  \Eprint {http://arxiv.org/abs/2302.03925} {arXiv:2302.03925
  [cond-mat.str-el]} \BibitemShut {NoStop}%
\bibitem [{\citenamefont {Catena}\ \emph {et~al.}(2021)\citenamefont {Catena},
  \citenamefont {Emken}, \citenamefont {Matas}, \citenamefont {Spaldin},\ and\
  \citenamefont {Urdshals}}]{Catena:2021qsr}%
  \BibitemOpen
  \bibfield  {author} {\bibinfo {author} {\bibfnamefont {Riccardo}\
  \bibnamefont {Catena}}, \bibinfo {author} {\bibfnamefont {Timon}\
  \bibnamefont {Emken}}, \bibinfo {author} {\bibfnamefont {Marek}\ \bibnamefont
  {Matas}}, \bibinfo {author} {\bibfnamefont {Nicola~A.}\ \bibnamefont
  {Spaldin}}, \ and\ \bibinfo {author} {\bibfnamefont {Einar}\ \bibnamefont
  {Urdshals}},\ }\bibfield  {title} {\enquote {\bibinfo {title} {{Crystal
  responses to general dark matter-electron interactions}},}\ }\href {\doibase
  10.1103/PhysRevResearch.3.033149} {\bibfield  {journal} {\bibinfo  {journal}
  {Phys. Rev. Res.}\ }\textbf {\bibinfo {volume} {3}},\ \bibinfo {pages}
  {033149} (\bibinfo {year} {2021})},\ \Eprint
  {http://arxiv.org/abs/2105.02233} {arXiv:2105.02233 [hep-ph]} \BibitemShut
  {NoStop}%
\bibitem [{\citenamefont {Catena}\ \emph {et~al.}(2019)\citenamefont {Catena},
  \citenamefont {Fridell},\ and\ \citenamefont {Krauss}}]{Catena:2019hzw}%
  \BibitemOpen
  \bibfield  {author} {\bibinfo {author} {\bibfnamefont {Riccardo}\
  \bibnamefont {Catena}}, \bibinfo {author} {\bibfnamefont {Kåre}\
  \bibnamefont {Fridell}}, \ and\ \bibinfo {author} {\bibfnamefont {Martin~B.}\
  \bibnamefont {Krauss}},\ }\bibfield  {title} {\enquote {\bibinfo {title}
  {{Non-relativistic Effective Interactions of Spin 1 Dark Matter}},}\ }\href
  {\doibase 10.1007/JHEP08(2019)030} {\bibfield  {journal} {\bibinfo  {journal}
  {JHEP}\ }\textbf {\bibinfo {volume} {08}},\ \bibinfo {pages} {030} (\bibinfo
  {year} {2019})},\ \Eprint {http://arxiv.org/abs/1907.02910} {arXiv:1907.02910
  [hep-ph]} \BibitemShut {NoStop}%
\bibitem [{\citenamefont {Fan}\ \emph {et~al.}(2010)\citenamefont {Fan},
  \citenamefont {Reece},\ and\ \citenamefont {Wang}}]{Fan:2010gt}%
  \BibitemOpen
  \bibfield  {author} {\bibinfo {author} {\bibfnamefont {JiJi}\ \bibnamefont
  {Fan}}, \bibinfo {author} {\bibfnamefont {Matthew}\ \bibnamefont {Reece}}, \
  and\ \bibinfo {author} {\bibfnamefont {Lian-Tao}\ \bibnamefont {Wang}},\
  }\bibfield  {title} {\enquote {\bibinfo {title} {{Non-relativistic effective
  theory of dark matter direct detection}},}\ }\href {\doibase
  10.1088/1475-7516/2010/11/042} {\bibfield  {journal} {\bibinfo  {journal}
  {JCAP}\ }\textbf {\bibinfo {volume} {11}},\ \bibinfo {pages} {042} (\bibinfo
  {year} {2010})},\ \Eprint {http://arxiv.org/abs/1008.1591} {arXiv:1008.1591
  [hep-ph]} \BibitemShut {NoStop}%
\bibitem [{\citenamefont {Fitzpatrick}\ \emph {et~al.}(2013)\citenamefont
  {Fitzpatrick}, \citenamefont {Haxton}, \citenamefont {Katz}, \citenamefont
  {Lubbers},\ and\ \citenamefont {Xu}}]{Fitzpatrick:2012ix}%
  \BibitemOpen
  \bibfield  {author} {\bibinfo {author} {\bibfnamefont {A.~Liam}\ \bibnamefont
  {Fitzpatrick}}, \bibinfo {author} {\bibfnamefont {Wick}\ \bibnamefont
  {Haxton}}, \bibinfo {author} {\bibfnamefont {Emanuel}\ \bibnamefont {Katz}},
  \bibinfo {author} {\bibfnamefont {Nicholas}\ \bibnamefont {Lubbers}}, \ and\
  \bibinfo {author} {\bibfnamefont {Yiming}\ \bibnamefont {Xu}},\ }\bibfield
  {title} {\enquote {\bibinfo {title} {{The Effective Field Theory of Dark
  Matter Direct Detection}},}\ }\href {\doibase 10.1088/1475-7516/2013/02/004}
  {\bibfield  {journal} {\bibinfo  {journal} {JCAP}\ }\textbf {\bibinfo
  {volume} {1302}},\ \bibinfo {pages} {004} (\bibinfo {year} {2013})},\ \Eprint
  {http://arxiv.org/abs/1203.3542} {arXiv:1203.3542 [hep-ph]} \BibitemShut
  {NoStop}%
\bibitem [{\citenamefont {Catena}\ \emph {et~al.}(2020)\citenamefont {Catena},
  \citenamefont {Emken}, \citenamefont {Spaldin},\ and\ \citenamefont
  {Tarantino}}]{Catena:2019gfa}%
  \BibitemOpen
  \bibfield  {author} {\bibinfo {author} {\bibfnamefont {Riccardo}\
  \bibnamefont {Catena}}, \bibinfo {author} {\bibfnamefont {Timon}\
  \bibnamefont {Emken}}, \bibinfo {author} {\bibfnamefont {Nicola~A.}\
  \bibnamefont {Spaldin}}, \ and\ \bibinfo {author} {\bibfnamefont {Walter}\
  \bibnamefont {Tarantino}},\ }\bibfield  {title} {\enquote {\bibinfo {title}
  {{Atomic responses to general dark matter-electron interactions}},}\ }\href
  {\doibase 10.1103/PhysRevResearch.2.033195} {\bibfield  {journal} {\bibinfo
  {journal} {Phys. Rev. Res.}\ }\textbf {\bibinfo {volume} {2}},\ \bibinfo
  {pages} {033195} (\bibinfo {year} {2020})},\ \Eprint
  {http://arxiv.org/abs/1912.08204} {arXiv:1912.08204 [hep-ph]} \BibitemShut
  {NoStop}%
\bibitem [{\citenamefont {Baxter}\ \emph {et~al.}(2021)\citenamefont {Baxter}
  \emph {et~al.}}]{Baxter:2021pqo}%
  \BibitemOpen
  \bibfield  {author} {\bibinfo {author} {\bibfnamefont {Daniel}\ \bibnamefont
  {Baxter}} \emph {et~al.},\ }\bibfield  {title} {\enquote {\bibinfo {title}
  {{Recommended conventions for reporting results from direct dark matter
  searches}},}\ }\href {\doibase 10.1140/epjc/s10052-021-09655-y} {\bibfield
  {journal} {\bibinfo  {journal} {Eur. Phys. J. C}\ }\textbf {\bibinfo {volume}
  {81}},\ \bibinfo {pages} {907} (\bibinfo {year} {2021})},\ \Eprint
  {http://arxiv.org/abs/2105.00599} {arXiv:2105.00599 [hep-ex]} \BibitemShut
  {NoStop}%
\bibitem [{\citenamefont {An}\ \emph {et~al.}(2013)\citenamefont {An},
  \citenamefont {Pospelov},\ and\ \citenamefont {Pradler}}]{An:2013yua}%
  \BibitemOpen
  \bibfield  {author} {\bibinfo {author} {\bibfnamefont {Haipeng}\ \bibnamefont
  {An}}, \bibinfo {author} {\bibfnamefont {Maxim}\ \bibnamefont {Pospelov}}, \
  and\ \bibinfo {author} {\bibfnamefont {Josef}\ \bibnamefont {Pradler}},\
  }\bibfield  {title} {\enquote {\bibinfo {title} {{Dark Matter Detectors as
  Dark Photon Helioscopes}},}\ }\href {\doibase 10.1103/PhysRevLett.111.041302}
  {\bibfield  {journal} {\bibinfo  {journal} {Phys. Rev. Lett.}\ }\textbf
  {\bibinfo {volume} {111}},\ \bibinfo {pages} {041302} (\bibinfo {year}
  {2013})},\ \Eprint {http://arxiv.org/abs/1304.3461} {arXiv:1304.3461
  [hep-ph]} \BibitemShut {NoStop}%
\bibitem [{\citenamefont {Knipp}(1989)}]{knipp1989surface}%
  \BibitemOpen
  \bibfield  {author} {\bibinfo {author} {\bibfnamefont {Peter}\ \bibnamefont
  {Knipp}},\ }\bibfield  {title} {\enquote {\bibinfo {title} {Surface phonons
  localized at step edges},}\ }\href {\doibase 10.1103/PhysRevB.40.7993}
  {\bibfield  {journal} {\bibinfo  {journal} {Phys. Rev. B}\ }\textbf {\bibinfo
  {volume} {40}},\ \bibinfo {pages} {7993} (\bibinfo {year}
  {1989})}\BibitemShut {NoStop}%
\bibitem [{\citenamefont {Luckyanova}\ \emph {et~al.}(2018)\citenamefont
  {Luckyanova}, \citenamefont {Mendoza}, \citenamefont {Lu}, \citenamefont
  {Song}, \citenamefont {Huang}, \citenamefont {Zhou}, \citenamefont {Li},
  \citenamefont {Dong}, \citenamefont {Zhou}, \citenamefont {Garlow} \emph
  {et~al.}}]{luckyanova2018phonon}%
  \BibitemOpen
  \bibfield  {author} {\bibinfo {author} {\bibfnamefont {Maria~N.}\
  \bibnamefont {Luckyanova}}, \bibinfo {author} {\bibfnamefont {Jonathan}\
  \bibnamefont {Mendoza}}, \bibinfo {author} {\bibfnamefont {Hong}\
  \bibnamefont {Lu}}, \bibinfo {author} {\bibfnamefont {Bai}\ \bibnamefont
  {Song}}, \bibinfo {author} {\bibfnamefont {Shengxi}\ \bibnamefont {Huang}},
  \bibinfo {author} {\bibfnamefont {Jiawei}\ \bibnamefont {Zhou}}, \bibinfo
  {author} {\bibfnamefont {Mingda}\ \bibnamefont {Li}}, \bibinfo {author}
  {\bibfnamefont {Yongqi}\ \bibnamefont {Dong}}, \bibinfo {author}
  {\bibfnamefont {Hua}\ \bibnamefont {Zhou}}, \bibinfo {author} {\bibfnamefont
  {Joseph}\ \bibnamefont {Garlow}},  \emph {et~al.},\ }\bibfield  {title}
  {\enquote {\bibinfo {title} {Phonon localization in heat conduction},}\
  }\href {\doibase 10.1126/sciadv.aat9460} {\bibfield  {journal} {\bibinfo
  {journal} {Sci. Adv.}\ }\textbf {\bibinfo {volume} {4}},\ \bibinfo {pages}
  {eaat9460} (\bibinfo {year} {2018})}\BibitemShut {NoStop}%
\bibitem [{\citenamefont {Vasyukov}\ \emph {et~al.}(2013)\citenamefont
  {Vasyukov}, \citenamefont {Anahory}, \citenamefont {Embon}, \citenamefont
  {Halbertal}, \citenamefont {Cuppens}, \citenamefont {Neeman}, \citenamefont
  {Finkler}, \citenamefont {Segev}, \citenamefont {Myasoedov}, \citenamefont
  {Rappaport} \emph {et~al.}}]{vasyukov2013scanning}%
  \BibitemOpen
  \bibfield  {author} {\bibinfo {author} {\bibfnamefont {Denis}\ \bibnamefont
  {Vasyukov}}, \bibinfo {author} {\bibfnamefont {Yonathan}\ \bibnamefont
  {Anahory}}, \bibinfo {author} {\bibfnamefont {Lior}\ \bibnamefont {Embon}},
  \bibinfo {author} {\bibfnamefont {Dorri}\ \bibnamefont {Halbertal}}, \bibinfo
  {author} {\bibfnamefont {Jo}~\bibnamefont {Cuppens}}, \bibinfo {author}
  {\bibfnamefont {Lior}\ \bibnamefont {Neeman}}, \bibinfo {author}
  {\bibfnamefont {Amit}\ \bibnamefont {Finkler}}, \bibinfo {author}
  {\bibfnamefont {Yehonathan}\ \bibnamefont {Segev}}, \bibinfo {author}
  {\bibfnamefont {Yuri}\ \bibnamefont {Myasoedov}}, \bibinfo {author}
  {\bibfnamefont {Michael~L}\ \bibnamefont {Rappaport}},  \emph {et~al.},\
  }\bibfield  {title} {\enquote {\bibinfo {title} {A scanning superconducting
  quantum interference device with single electron spin sensitivity},}\ }\href
  {\doibase 10.1038/nnano.2013.169} {\bibfield  {journal} {\bibinfo  {journal}
  {Nat. Nanotechnol.}\ }\textbf {\bibinfo {volume} {8}},\ \bibinfo {pages}
  {639--644} (\bibinfo {year} {2013})}\BibitemShut {NoStop}%
\bibitem [{\citenamefont {Grinolds}\ \emph {et~al.}(2013)\citenamefont
  {Grinolds}, \citenamefont {Hong}, \citenamefont {Maletinsky}, \citenamefont
  {Luan}, \citenamefont {Lukin}, \citenamefont {Walsworth},\ and\ \citenamefont
  {Yacoby}}]{grinolds2013nanoscale}%
  \BibitemOpen
  \bibfield  {author} {\bibinfo {author} {\bibfnamefont {Michael~Sean}\
  \bibnamefont {Grinolds}}, \bibinfo {author} {\bibfnamefont {Sungkun}\
  \bibnamefont {Hong}}, \bibinfo {author} {\bibfnamefont {Patrick}\
  \bibnamefont {Maletinsky}}, \bibinfo {author} {\bibfnamefont {Lan}\
  \bibnamefont {Luan}}, \bibinfo {author} {\bibfnamefont {Mikhail~D.}\
  \bibnamefont {Lukin}}, \bibinfo {author} {\bibfnamefont {Ronald~Lee}\
  \bibnamefont {Walsworth}}, \ and\ \bibinfo {author} {\bibfnamefont {Amir}\
  \bibnamefont {Yacoby}},\ }\bibfield  {title} {\enquote {\bibinfo {title}
  {Nanoscale magnetic imaging of a single electron spin under ambient
  conditions},}\ }\href {\doibase 10.1038/nphys2543} {\bibfield  {journal}
  {\bibinfo  {journal} {Nat. Phys.}\ }\textbf {\bibinfo {volume} {9}},\
  \bibinfo {pages} {215--219} (\bibinfo {year} {2013})}\BibitemShut {NoStop}%
\bibitem [{\citenamefont {Steinert}\ \emph {et~al.}(2010)\citenamefont
  {Steinert}, \citenamefont {Dolde}, \citenamefont {Neumann}, \citenamefont
  {Aird}, \citenamefont {Naydenov}, \citenamefont {Balasubramanian},
  \citenamefont {Jelezko},\ and\ \citenamefont {Wrachtrup}}]{steinert2010high}%
  \BibitemOpen
  \bibfield  {author} {\bibinfo {author} {\bibfnamefont {Steffen}\ \bibnamefont
  {Steinert}}, \bibinfo {author} {\bibfnamefont {Florian}\ \bibnamefont
  {Dolde}}, \bibinfo {author} {\bibfnamefont {Philipp}\ \bibnamefont
  {Neumann}}, \bibinfo {author} {\bibfnamefont {Andrew}\ \bibnamefont {Aird}},
  \bibinfo {author} {\bibfnamefont {Boris}\ \bibnamefont {Naydenov}}, \bibinfo
  {author} {\bibfnamefont {Gopalakrishnan}\ \bibnamefont {Balasubramanian}},
  \bibinfo {author} {\bibfnamefont {Fedor}\ \bibnamefont {Jelezko}}, \ and\
  \bibinfo {author} {\bibfnamefont {Joerg}\ \bibnamefont {Wrachtrup}},\
  }\bibfield  {title} {\enquote {\bibinfo {title} {High sensitivity magnetic
  imaging using an array of spins in diamond},}\ }\href {\doibase
  10.1063/1.3385689} {\bibfield  {journal} {\bibinfo  {journal} {Rev. Sci.
  Instrum.}\ }\textbf {\bibinfo {volume} {81}},\ \bibinfo {pages} {043705}
  (\bibinfo {year} {2010})}\BibitemShut {NoStop}%
\bibitem [{\citenamefont {Drung}\ \emph {et~al.}(2007)\citenamefont {Drung}
  \emph {et~al.}}]{drung2007highly}%
  \BibitemOpen
  \bibfield  {author} {\bibinfo {author} {\bibfnamefont {Dietmar}\ \bibnamefont
  {Drung}} \emph {et~al.},\ }\bibfield  {title} {\enquote {\bibinfo {title}
  {Highly sensitive and easy-to-use {SQUID} sensors},}\ }\href {\doibase
  10.1109/TASC.2007.897403} {\bibfield  {journal} {\bibinfo  {journal} {IEEE
  Trans. Appl. Supercond.}\ }\textbf {\bibinfo {volume} {17}},\ \bibinfo
  {pages} {699--704} (\bibinfo {year} {2007})}\BibitemShut {NoStop}%
\bibitem [{\citenamefont {Carney}\ \emph {et~al.}(2021)\citenamefont {Carney},
  \citenamefont {Hook}, \citenamefont {Liu}, \citenamefont {Taylor},\ and\
  \citenamefont {Zhao}}]{carney2021ultralight}%
  \BibitemOpen
  \bibfield  {author} {\bibinfo {author} {\bibfnamefont {Daniel}\ \bibnamefont
  {Carney}}, \bibinfo {author} {\bibfnamefont {Anson}\ \bibnamefont {Hook}},
  \bibinfo {author} {\bibfnamefont {Zhen}\ \bibnamefont {Liu}}, \bibinfo
  {author} {\bibfnamefont {Jacob~M.}\ \bibnamefont {Taylor}}, \ and\ \bibinfo
  {author} {\bibfnamefont {Yue}\ \bibnamefont {Zhao}},\ }\bibfield  {title}
  {\enquote {\bibinfo {title} {Ultralight dark matter detection with mechanical
  quantum sensors},}\ }\href {\doibase 10.1088/1367-2630/abd9e7} {\bibfield
  {journal} {\bibinfo  {journal} {New J. Phys.}\ }\textbf {\bibinfo {volume}
  {23}},\ \bibinfo {pages} {023041} (\bibinfo {year} {2021})}\BibitemShut
  {NoStop}%
\bibitem [{\citenamefont {MacCabe}\ \emph {et~al.}(2020)\citenamefont
  {MacCabe}, \citenamefont {Ren}, \citenamefont {Luo}, \citenamefont {Cohen},
  \citenamefont {Zhou}, \citenamefont {Sipahigil}, \citenamefont
  {Mirhosseini},\ and\ \citenamefont {Painter}}]{maccabe2020nano}%
  \BibitemOpen
  \bibfield  {author} {\bibinfo {author} {\bibfnamefont {Gregory~S.}\
  \bibnamefont {MacCabe}}, \bibinfo {author} {\bibfnamefont {Hengjiang}\
  \bibnamefont {Ren}}, \bibinfo {author} {\bibfnamefont {Jie}\ \bibnamefont
  {Luo}}, \bibinfo {author} {\bibfnamefont {Justin~D.}\ \bibnamefont {Cohen}},
  \bibinfo {author} {\bibfnamefont {Hengyun}\ \bibnamefont {Zhou}}, \bibinfo
  {author} {\bibfnamefont {Alp}\ \bibnamefont {Sipahigil}}, \bibinfo {author}
  {\bibfnamefont {Mohammad}\ \bibnamefont {Mirhosseini}}, \ and\ \bibinfo
  {author} {\bibfnamefont {Oskar}\ \bibnamefont {Painter}},\ }\bibfield
  {title} {\enquote {\bibinfo {title} {Nano-acoustic resonator with ultralong
  phonon lifetime},}\ }\href {\doibase 10.1126/science.abc7312} {\bibfield
  {journal} {\bibinfo  {journal} {Science}\ }\textbf {\bibinfo {volume}
  {370}},\ \bibinfo {pages} {840--843} (\bibinfo {year} {2020})}\BibitemShut
  {NoStop}%
\bibitem [{\citenamefont {Pandey}\ \emph {et~al.}(2018)\citenamefont {Pandey},
  \citenamefont {Polanco}, \citenamefont {Cooper}, \citenamefont {Parker},\
  and\ \citenamefont {Lindsay}}]{pandey2018symmetry}%
  \BibitemOpen
  \bibfield  {author} {\bibinfo {author} {\bibfnamefont {Tribhuwan}\
  \bibnamefont {Pandey}}, \bibinfo {author} {\bibfnamefont {Carlos~A.}\
  \bibnamefont {Polanco}}, \bibinfo {author} {\bibfnamefont {Valentino~R.}\
  \bibnamefont {Cooper}}, \bibinfo {author} {\bibfnamefont {David~S.}\
  \bibnamefont {Parker}}, \ and\ \bibinfo {author} {\bibfnamefont {Lucas}\
  \bibnamefont {Lindsay}},\ }\bibfield  {title} {\enquote {\bibinfo {title}
  {Symmetry-driven phonon chirality and transport in one-dimensional and bulk
  {B}a$_3${N}-derived materials},}\ }\href {\doibase
  10.1103/PhysRevB.98.241405} {\bibfield  {journal} {\bibinfo  {journal} {Phys.
  Rev. B}\ }\textbf {\bibinfo {volume} {98}},\ \bibinfo {pages} {241405}
  (\bibinfo {year} {2018})}\BibitemShut {NoStop}%
\bibitem [{\citenamefont {Chaudhary}\ \emph {et~al.}(2023)\citenamefont
  {Chaudhary}, \citenamefont {Juraschek}, \citenamefont {Rodriguez-Vega},\ and\
  \citenamefont {Fiete}}]{chaudhary2023giant}%
  \BibitemOpen
  \bibfield  {author} {\bibinfo {author} {\bibfnamefont {Swati}\ \bibnamefont
  {Chaudhary}}, \bibinfo {author} {\bibfnamefont {Dominik~M.}\ \bibnamefont
  {Juraschek}}, \bibinfo {author} {\bibfnamefont {Martin}\ \bibnamefont
  {Rodriguez-Vega}}, \ and\ \bibinfo {author} {\bibfnamefont {Gregory~A.}\
  \bibnamefont {Fiete}},\ }\bibfield  {title} {\enquote {\bibinfo {title}
  {Giant effective magnetic moments of chiral phonons from orbit-lattice
  coupling},}\ }\href {\doibase 10.48550/arXiv.2306.11630} {\  (\bibinfo {year}
  {2023}),\ 10.48550/arXiv.2306.11630},\ \Eprint
  {http://arxiv.org/abs/2306.11630} {arXiv:2306.11630 [cond-mat.mes-hall]}
  \BibitemShut {NoStop}%
\bibitem [{\citenamefont {Wang}\ \emph {et~al.}(2012)\citenamefont {Wang},
  \citenamefont {Zhang},\ and\ \citenamefont {Lin}}]{wang2012rational}%
  \BibitemOpen
  \bibfield  {author} {\bibinfo {author} {\bibfnamefont {Cheng}\ \bibnamefont
  {Wang}}, \bibinfo {author} {\bibfnamefont {Teng}\ \bibnamefont {Zhang}}, \
  and\ \bibinfo {author} {\bibfnamefont {Wenbin}\ \bibnamefont {Lin}},\
  }\bibfield  {title} {\enquote {\bibinfo {title} {Rational synthesis of
  noncentrosymmetric metal--organic frameworks for second-order nonlinear
  optics},}\ }\href {\doibase 10.1021/cr200252n} {\bibfield  {journal}
  {\bibinfo  {journal} {Chem. Rev.}\ }\textbf {\bibinfo {volume} {112}},\
  \bibinfo {pages} {1084--1104} (\bibinfo {year} {2012})}\BibitemShut {NoStop}%
\bibitem [{\citenamefont {Gonzalez-Nelson}\ \emph {et~al.}(2019)\citenamefont
  {Gonzalez-Nelson}, \citenamefont {Coudert},\ and\ \citenamefont {van
  Der~Veen}}]{gonzalez2019rotational}%
  \BibitemOpen
  \bibfield  {author} {\bibinfo {author} {\bibfnamefont {Adrian}\ \bibnamefont
  {Gonzalez-Nelson}}, \bibinfo {author} {\bibfnamefont {Fran{\c{c}}ois-Xavier}\
  \bibnamefont {Coudert}}, \ and\ \bibinfo {author} {\bibfnamefont
  {Monique~A.}\ \bibnamefont {van Der~Veen}},\ }\bibfield  {title} {\enquote
  {\bibinfo {title} {Rotational dynamics of linkers in metal--organic
  frameworks},}\ }\href {\doibase 10.3390/nano9030330} {\bibfield  {journal}
  {\bibinfo  {journal} {Nanomater.}\ }\textbf {\bibinfo {volume} {9}},\
  \bibinfo {pages} {330} (\bibinfo {year} {2019})}\BibitemShut {NoStop}%
\bibitem [{\citenamefont {Das}\ \emph {et~al.}(2017)\citenamefont {Das},
  \citenamefont {Chatterjee}, \citenamefont {Bhunia}, \citenamefont {Mondal},
  \citenamefont {Mitra}, \citenamefont {Kumari}, \citenamefont {Pradhan},\ and\
  \citenamefont {Bhaumik}}]{das2017new}%
  \BibitemOpen
  \bibfield  {author} {\bibinfo {author} {\bibfnamefont {Sabuj~Kanti}\
  \bibnamefont {Das}}, \bibinfo {author} {\bibfnamefont {Sauvik}\ \bibnamefont
  {Chatterjee}}, \bibinfo {author} {\bibfnamefont {Subhajit}\ \bibnamefont
  {Bhunia}}, \bibinfo {author} {\bibfnamefont {Abhishake}\ \bibnamefont
  {Mondal}}, \bibinfo {author} {\bibfnamefont {Partha}\ \bibnamefont {Mitra}},
  \bibinfo {author} {\bibfnamefont {Vandana}\ \bibnamefont {Kumari}}, \bibinfo
  {author} {\bibfnamefont {Anirban}\ \bibnamefont {Pradhan}}, \ and\ \bibinfo
  {author} {\bibfnamefont {Asim}\ \bibnamefont {Bhaumik}},\ }\bibfield  {title}
  {\enquote {\bibinfo {title} {A new strongly paramagnetic cerium-containing
  microporous {MOF} for {CO$_2$} fixation under ambient conditions},}\ }\href
  {\doibase 10.1039/C7DT02040F} {\bibfield  {journal} {\bibinfo  {journal}
  {Dalton Trans.}\ }\textbf {\bibinfo {volume} {46}},\ \bibinfo {pages}
  {13783--13792} (\bibinfo {year} {2017})}\BibitemShut {NoStop}%
\bibitem [{\citenamefont {Deng}\ \emph {et~al.}(2021)\citenamefont {Deng},
  \citenamefont {Shi}, \citenamefont {Wong}, \citenamefont {Wu}, \citenamefont
  {Yang}, \citenamefont {Zheng}, \citenamefont {Pan},\ and\ \citenamefont
  {Yang}}]{deng2021designing}%
  \BibitemOpen
  \bibfield  {author} {\bibinfo {author} {\bibfnamefont {Tianqi}\ \bibnamefont
  {Deng}}, \bibinfo {author} {\bibfnamefont {Wen}\ \bibnamefont {Shi}},
  \bibinfo {author} {\bibfnamefont {Zicong~Marvin}\ \bibnamefont {Wong}},
  \bibinfo {author} {\bibfnamefont {Gang}\ \bibnamefont {Wu}}, \bibinfo
  {author} {\bibfnamefont {Xiaoping}\ \bibnamefont {Yang}}, \bibinfo {author}
  {\bibfnamefont {Jin-Cheng}\ \bibnamefont {Zheng}}, \bibinfo {author}
  {\bibfnamefont {Hui}\ \bibnamefont {Pan}}, \ and\ \bibinfo {author}
  {\bibfnamefont {Shuo-Wang}\ \bibnamefont {Yang}},\ }\bibfield  {title}
  {\enquote {\bibinfo {title} {Designing intrinsic topological insulators in
  two-dimensional metal--organic frameworks},}\ }\href {\doibase
  10.1021/acs.jpclett.1c01731} {\bibfield  {journal} {\bibinfo  {journal} {J.
  Phys. Chem. Lett.}\ }\textbf {\bibinfo {volume} {12}},\ \bibinfo {pages}
  {6934--6940} (\bibinfo {year} {2021})}\BibitemShut {NoStop}%
\bibitem [{\citenamefont {Ni}\ \emph {et~al.}(2022)\citenamefont {Ni},
  \citenamefont {Huang},\ and\ \citenamefont {Br{\'e}das}}]{ni2022emergence}%
  \BibitemOpen
  \bibfield  {author} {\bibinfo {author} {\bibfnamefont {Xiaojuan}\
  \bibnamefont {Ni}}, \bibinfo {author} {\bibfnamefont {Huaqing}\ \bibnamefont
  {Huang}}, \ and\ \bibinfo {author} {\bibfnamefont {Jean-Luc}\ \bibnamefont
  {Br{\'e}das}},\ }\bibfield  {title} {\enquote {\bibinfo {title} {Emergence of
  a two-dimensional topological {D}irac semimetal phase in a
  phthalocyanine-based covalent organic framework},}\ }\href {\doibase
  10.1021/acs.chemmater.1c04317} {\bibfield  {journal} {\bibinfo  {journal}
  {Chem. Mater.}\ }\textbf {\bibinfo {volume} {34}},\ \bibinfo {pages}
  {3178--3184} (\bibinfo {year} {2022})}\BibitemShut {NoStop}%
\bibitem [{\citenamefont {Kim}\ \emph {et~al.}(2019)\citenamefont {Kim},
  \citenamefont {Park}, \citenamefont {Paczesny},\ and\ \citenamefont
  {Grzybowski}}]{kim2019uniform}%
  \BibitemOpen
  \bibfield  {author} {\bibinfo {author} {\bibfnamefont {Namhun}\ \bibnamefont
  {Kim}}, \bibinfo {author} {\bibfnamefont {Jun~Heuk}\ \bibnamefont {Park}},
  \bibinfo {author} {\bibfnamefont {Jan}\ \bibnamefont {Paczesny}}, \ and\
  \bibinfo {author} {\bibfnamefont {Bartosz~A.}\ \bibnamefont {Grzybowski}},\
  }\bibfield  {title} {\enquote {\bibinfo {title} {Uniform and directional
  growth of centimeter-sized single crystals of cyclodextrin-based metal
  organic frameworks},}\ }\href {\doibase 10.1039/C9CE00026G} {\bibfield
  {journal} {\bibinfo  {journal} {CrystEngComm}\ }\textbf {\bibinfo {volume}
  {21}},\ \bibinfo {pages} {1867--1871} (\bibinfo {year} {2019})}\BibitemShut
  {NoStop}%
\bibitem [{\citenamefont {Kim}\ \emph {et~al.}(2020)\citenamefont {Kim},
  \citenamefont {Ellis}, \citenamefont {Howard},\ and\ \citenamefont
  {Ohodnicki}}]{kim2020centimeter}%
  \BibitemOpen
  \bibfield  {author} {\bibinfo {author} {\bibfnamefont {Ki-Joong}\
  \bibnamefont {Kim}}, \bibinfo {author} {\bibfnamefont {James~E.}\
  \bibnamefont {Ellis}}, \bibinfo {author} {\bibfnamefont {Bret~H.}\
  \bibnamefont {Howard}}, \ and\ \bibinfo {author} {\bibfnamefont {Paul~R.}\
  \bibnamefont {Ohodnicki}},\ }\bibfield  {title} {\enquote {\bibinfo {title}
  {Centimeter-scale pillared-layer metal--organic framework thin films mediated
  by hydroxy double salt intermediates for {CO$_2$} sensor applications},}\
  }\href {\doibase 10.1021/acsami.0c19621} {\bibfield  {journal} {\bibinfo
  {journal} {ACS Appl. Mater. Interfaces}\ }\textbf {\bibinfo {volume} {13}},\
  \bibinfo {pages} {2062--2071} (\bibinfo {year} {2020})}\BibitemShut {NoStop}%
\bibitem [{\citenamefont {Gonze}\ \emph {et~al.}(2020)\citenamefont {Gonze},
  \citenamefont {Amadon}, \citenamefont {Antonius}, \citenamefont {Arnardi},
  \citenamefont {Baguet}, \citenamefont {Beuken}, \citenamefont {Bieder},
  \citenamefont {Bottin}, \citenamefont {Bouchet}, \citenamefont {Bousquet}
  \emph {et~al.}}]{gonze2020abinit}%
  \BibitemOpen
  \bibfield  {author} {\bibinfo {author} {\bibfnamefont {Xavier}\ \bibnamefont
  {Gonze}}, \bibinfo {author} {\bibfnamefont {Bernard}\ \bibnamefont {Amadon}},
  \bibinfo {author} {\bibfnamefont {Gabriel}\ \bibnamefont {Antonius}},
  \bibinfo {author} {\bibfnamefont {Fr{\'e}d{\'e}ric}\ \bibnamefont {Arnardi}},
  \bibinfo {author} {\bibfnamefont {Lucas}\ \bibnamefont {Baguet}}, \bibinfo
  {author} {\bibfnamefont {Jean-Michel}\ \bibnamefont {Beuken}}, \bibinfo
  {author} {\bibfnamefont {Jordan}\ \bibnamefont {Bieder}}, \bibinfo {author}
  {\bibfnamefont {Fran{\c{c}}ois}\ \bibnamefont {Bottin}}, \bibinfo {author}
  {\bibfnamefont {Johann}\ \bibnamefont {Bouchet}}, \bibinfo {author}
  {\bibfnamefont {Eric}\ \bibnamefont {Bousquet}},  \emph {et~al.},\ }\bibfield
   {title} {\enquote {\bibinfo {title} {The {A\sc{binit}} project: Impact,
  environment and recent developments},}\ }\href {\doibase
  10.1016/j.cpc.2019.107042} {\bibfield  {journal} {\bibinfo  {journal}
  {Comput. Phys. Commun.}\ }\textbf {\bibinfo {volume} {248}},\ \bibinfo
  {pages} {107042} (\bibinfo {year} {2020})}\BibitemShut {NoStop}%
\bibitem [{\citenamefont {Gonze}\ and\ \citenamefont
  {Lee}(1997)}]{gonze1997dynamical}%
  \BibitemOpen
  \bibfield  {author} {\bibinfo {author} {\bibfnamefont {Xavier}\ \bibnamefont
  {Gonze}}\ and\ \bibinfo {author} {\bibfnamefont {Changyol}\ \bibnamefont
  {Lee}},\ }\bibfield  {title} {\enquote {\bibinfo {title} {Dynamical matrices,
  {B}orn effective charges, dielectric permittivity tensors, and interatomic
  force constants from density-functional perturbation theory},}\ }\href
  {\doibase 10.1103/PhysRevB.55.10355} {\bibfield  {journal} {\bibinfo
  {journal} {Phys. Rev. B}\ }\textbf {\bibinfo {volume} {55}},\ \bibinfo
  {pages} {10355} (\bibinfo {year} {1997})}\BibitemShut {NoStop}%
\bibitem [{\citenamefont {Bottin}\ \emph {et~al.}(2008)\citenamefont {Bottin},
  \citenamefont {Leroux}, \citenamefont {Knyazev},\ and\ \citenamefont
  {Z{\'e}rah}}]{bottin2008large}%
  \BibitemOpen
  \bibfield  {author} {\bibinfo {author} {\bibfnamefont {Fran{\c{c}}ois}\
  \bibnamefont {Bottin}}, \bibinfo {author} {\bibfnamefont {St{\'e}phane}\
  \bibnamefont {Leroux}}, \bibinfo {author} {\bibfnamefont {Andrew}\
  \bibnamefont {Knyazev}}, \ and\ \bibinfo {author} {\bibfnamefont {Gilles}\
  \bibnamefont {Z{\'e}rah}},\ }\bibfield  {title} {\enquote {\bibinfo {title}
  {Large-scale ab initio calculations based on three levels of
  parallelization},}\ }\href {\doibase 10.1016/j.commatsci.2007.07.019}
  {\bibfield  {journal} {\bibinfo  {journal} {Comput. Mater. Sci}\ }\textbf
  {\bibinfo {volume} {42}},\ \bibinfo {pages} {329--336} (\bibinfo {year}
  {2008})}\BibitemShut {NoStop}%
\bibitem [{\citenamefont {Bj{\"o}rkman}(2011)}]{bjorkman2011cif2cell}%
  \BibitemOpen
  \bibfield  {author} {\bibinfo {author} {\bibfnamefont {Torbj{\"o}rn}\
  \bibnamefont {Bj{\"o}rkman}},\ }\bibfield  {title} {\enquote {\bibinfo
  {title} {{CIF}2cell: generating geometries for electronic structure
  programs},}\ }\href {\doibase 10.1016/j.cpc.2011.01.013} {\bibfield
  {journal} {\bibinfo  {journal} {Comput. Phys. Commun.}\ }\textbf {\bibinfo
  {volume} {182}},\ \bibinfo {pages} {1183--1186} (\bibinfo {year}
  {2011})}\BibitemShut {NoStop}%
\bibitem [{\citenamefont {Perdew}\ \emph {et~al.}(1996)\citenamefont {Perdew},
  \citenamefont {Burke},\ and\ \citenamefont
  {Ernzerhof}}]{perdew1996generalized}%
  \BibitemOpen
  \bibfield  {author} {\bibinfo {author} {\bibfnamefont {John~P.}\ \bibnamefont
  {Perdew}}, \bibinfo {author} {\bibfnamefont {Kieron}\ \bibnamefont {Burke}},
  \ and\ \bibinfo {author} {\bibfnamefont {Matthias}\ \bibnamefont
  {Ernzerhof}},\ }\bibfield  {title} {\enquote {\bibinfo {title} {Generalized
  gradient approximation made simple},}\ }\href {\doibase
  10.1103/PhysRevLett.77.3865} {\bibfield  {journal} {\bibinfo  {journal}
  {Phys. Rev. Lett.}\ }\textbf {\bibinfo {volume} {77}},\ \bibinfo {pages}
  {3865} (\bibinfo {year} {1996})}\BibitemShut {NoStop}%
\bibitem [{\citenamefont {Monkhorst}\ and\ \citenamefont
  {Pack}(1976)}]{monkhorst1976special}%
  \BibitemOpen
  \bibfield  {author} {\bibinfo {author} {\bibfnamefont {Hendrik~J.}\
  \bibnamefont {Monkhorst}}\ and\ \bibinfo {author} {\bibfnamefont {James~D.}\
  \bibnamefont {Pack}},\ }\bibfield  {title} {\enquote {\bibinfo {title}
  {Special points for {B}rillouin-zone integrations},}\ }\href {\doibase
  10.1103/PhysRevB.13.5188} {\bibfield  {journal} {\bibinfo  {journal} {Phys.
  Rev. B}\ }\textbf {\bibinfo {volume} {13}},\ \bibinfo {pages} {5188}
  (\bibinfo {year} {1976})}\BibitemShut {NoStop}%
\bibitem [{\citenamefont {Grimme}\ \emph {et~al.}(2010)\citenamefont {Grimme},
  \citenamefont {Antony}, \citenamefont {Ehrlich},\ and\ \citenamefont
  {Krieg}}]{grimme2010consistent}%
  \BibitemOpen
  \bibfield  {author} {\bibinfo {author} {\bibfnamefont {Stefan}\ \bibnamefont
  {Grimme}}, \bibinfo {author} {\bibfnamefont {Jens}\ \bibnamefont {Antony}},
  \bibinfo {author} {\bibfnamefont {Stephan}\ \bibnamefont {Ehrlich}}, \ and\
  \bibinfo {author} {\bibfnamefont {Helge}\ \bibnamefont {Krieg}},\ }\bibfield
  {title} {\enquote {\bibinfo {title} {A consistent and accurate \textit{ab
  initio} parametrization of density functional dispersion correction ({DFT-D})
  for the 94 elements {H}--{P}u},}\ }\href {\doibase 10.1063/1.3382344}
  {\bibfield  {journal} {\bibinfo  {journal} {J. Chem. Phys.}\ }\textbf
  {\bibinfo {volume} {132}},\ \bibinfo {pages} {154104} (\bibinfo {year}
  {2010})}\BibitemShut {NoStop}%
\bibitem [{\citenamefont {Petrosyants}\ and\ \citenamefont
  {Ilyukhin}(2010)}]{petrosyants2010organometallic}%
  \BibitemOpen
  \bibfield  {author} {\bibinfo {author} {\bibfnamefont {Svetlana~P.}\
  \bibnamefont {Petrosyants}}\ and\ \bibinfo {author} {\bibfnamefont
  {Andrei~B.}\ \bibnamefont {Ilyukhin}},\ }\bibfield  {title} {\enquote
  {\bibinfo {title} {Organometallic polymers {MF$_3$}(4,$4'$-{B}ipy) with {M} =
  {G}a and {I}n},}\ }\href {\doibase 10.1134/S0036023610010079} {\bibfield
  {journal} {\bibinfo  {journal} {Russ. J. Inorg. Chem.}\ }\textbf {\bibinfo
  {volume} {55}},\ \bibinfo {pages} {30--33} (\bibinfo {year}
  {2010})}\BibitemShut {NoStop}%
\bibitem [{\citenamefont {Hinuma}\ \emph {et~al.}(2017)\citenamefont {Hinuma},
  \citenamefont {Pizzi}, \citenamefont {Kumagai}, \citenamefont {Oba},\ and\
  \citenamefont {Tanaka}}]{hinuma2017band}%
  \BibitemOpen
  \bibfield  {author} {\bibinfo {author} {\bibfnamefont {Yoyo}\ \bibnamefont
  {Hinuma}}, \bibinfo {author} {\bibfnamefont {Giovanni}\ \bibnamefont
  {Pizzi}}, \bibinfo {author} {\bibfnamefont {Yu}~\bibnamefont {Kumagai}},
  \bibinfo {author} {\bibfnamefont {Fumiyasu}\ \bibnamefont {Oba}}, \ and\
  \bibinfo {author} {\bibfnamefont {Isao}\ \bibnamefont {Tanaka}},\ }\bibfield
  {title} {\enquote {\bibinfo {title} {Band structure diagram paths based on
  crystallography},}\ }\href {\doibase 10.1016/j.commatsci.2016.10.015}
  {\bibfield  {journal} {\bibinfo  {journal} {Comput. Mater. Sci}\ }\textbf
  {\bibinfo {volume} {128}},\ \bibinfo {pages} {140--184} (\bibinfo {year}
  {2017})}\BibitemShut {NoStop}%
\bibitem [{\citenamefont {Kamencek}\ \emph {et~al.}(2019)\citenamefont
  {Kamencek}, \citenamefont {Bedoya-Mart{\'\i}nez},\ and\ \citenamefont
  {Zojer}}]{kamencek2019understanding}%
  \BibitemOpen
  \bibfield  {author} {\bibinfo {author} {\bibfnamefont {Tomas}\ \bibnamefont
  {Kamencek}}, \bibinfo {author} {\bibfnamefont {Natalia}\ \bibnamefont
  {Bedoya-Mart{\'\i}nez}}, \ and\ \bibinfo {author} {\bibfnamefont {Egbert}\
  \bibnamefont {Zojer}},\ }\bibfield  {title} {\enquote {\bibinfo {title}
  {Understanding phonon properties in isoreticular metal-organic frameworks
  from first principles},}\ }\href {\doibase 10.1103/PhysRevMaterials.3.116003}
  {\bibfield  {journal} {\bibinfo  {journal} {Phys. Rev. Mater.}\ }\textbf
  {\bibinfo {volume} {3}},\ \bibinfo {pages} {116003} (\bibinfo {year}
  {2019})}\BibitemShut {NoStop}%
\bibitem [{\citenamefont {Squires}(2012)}]{squires2012introduction}%
  \BibitemOpen
  \bibfield  {author} {\bibinfo {author} {\bibfnamefont {G.~L.}\ \bibnamefont
  {Squires}},\ }\href {\doibase 10.1017/CBO9781139107808} {\emph {\bibinfo
  {title} {Introduction to the Theory of Thermal Neutron Scattering}}}\
  (\bibinfo  {publisher} {Cambridge University Press},\ \bibinfo {year}
  {2012})\BibitemShut {NoStop}%
\bibitem [{\citenamefont {Essig}\ \emph {et~al.}(2016)\citenamefont {Essig},
  \citenamefont {Fernandez-Serra}, \citenamefont {Mardon}, \citenamefont
  {Soto}, \citenamefont {Volansky},\ and\ \citenamefont {Yu}}]{Essig:2015cda}%
  \BibitemOpen
  \bibfield  {author} {\bibinfo {author} {\bibfnamefont {Rouven}\ \bibnamefont
  {Essig}}, \bibinfo {author} {\bibfnamefont {Marivi}\ \bibnamefont
  {Fernandez-Serra}}, \bibinfo {author} {\bibfnamefont {Jeremy}\ \bibnamefont
  {Mardon}}, \bibinfo {author} {\bibfnamefont {Adrian}\ \bibnamefont {Soto}},
  \bibinfo {author} {\bibfnamefont {Tomer}\ \bibnamefont {Volansky}}, \ and\
  \bibinfo {author} {\bibfnamefont {Tien-Tien}\ \bibnamefont {Yu}},\ }\bibfield
   {title} {\enquote {\bibinfo {title} {{Direct Detection of sub-GeV Dark
  Matter with Semiconductor Targets}},}\ }\href {\doibase
  10.1007/JHEP05(2016)046} {\bibfield  {journal} {\bibinfo  {journal} {JHEP}\
  }\textbf {\bibinfo {volume} {05}},\ \bibinfo {pages} {046} (\bibinfo {year}
  {2016})},\ \Eprint {http://arxiv.org/abs/1509.01598} {arXiv:1509.01598
  [hep-ph]} \BibitemShut {NoStop}%
\bibitem [{\citenamefont {Abramoff}\ \emph {et~al.}(2019)\citenamefont
  {Abramoff} \emph {et~al.}}]{SENSEI:2019ibb}%
  \BibitemOpen
  \bibfield  {author} {\bibinfo {author} {\bibfnamefont {Orr}\ \bibnamefont
  {Abramoff}} \emph {et~al.} (\bibinfo {collaboration} {SENSEI}),\ }\bibfield
  {title} {\enquote {\bibinfo {title} {{SENSEI: Direct-Detection Constraints on
  Sub-GeV Dark Matter from a Shallow Underground Run Using a Prototype
  Skipper-CCD}},}\ }\href {\doibase 10.1103/PhysRevLett.122.161801} {\bibfield
  {journal} {\bibinfo  {journal} {Phys. Rev. Lett.}\ }\textbf {\bibinfo
  {volume} {122}},\ \bibinfo {pages} {161801} (\bibinfo {year} {2019})},\
  \Eprint {http://arxiv.org/abs/1901.10478} {arXiv:1901.10478 [hep-ex]}
  \BibitemShut {NoStop}%
\bibitem [{\citenamefont {Agnes}\ \emph {et~al.}(2018)\citenamefont {Agnes}
  \emph {et~al.}}]{DarkSide:2018ppu}%
  \BibitemOpen
  \bibfield  {author} {\bibinfo {author} {\bibfnamefont {Paolo}\ \bibnamefont
  {Agnes}} \emph {et~al.} (\bibinfo {collaboration} {DarkSide}),\ }\bibfield
  {title} {\enquote {\bibinfo {title} {{Constraints on Sub-GeV
  Dark-Matter\textendash{}Electron Scattering from the DarkSide-50
  Experiment}},}\ }\href {\doibase 10.1103/PhysRevLett.121.111303} {\bibfield
  {journal} {\bibinfo  {journal} {Phys. Rev. Lett.}\ }\textbf {\bibinfo
  {volume} {121}},\ \bibinfo {pages} {111303} (\bibinfo {year} {2018})},\
  \Eprint {http://arxiv.org/abs/1802.06998} {arXiv:1802.06998 [astro-ph.CO]}
  \BibitemShut {NoStop}%
\bibitem [{\citenamefont {Aguilar-Arevalo}\ \emph {et~al.}(2019)\citenamefont
  {Aguilar-Arevalo} \emph {et~al.}}]{DAMIC:2019dcn}%
  \BibitemOpen
  \bibfield  {author} {\bibinfo {author} {\bibfnamefont {Alexis}\ \bibnamefont
  {Aguilar-Arevalo}} \emph {et~al.} (\bibinfo {collaboration} {DAMIC}),\
  }\bibfield  {title} {\enquote {\bibinfo {title} {{Constraints on Light Dark
  Matter Particles Interacting with Electrons from DAMIC at SNOLAB}},}\ }\href
  {\doibase 10.1103/PhysRevLett.123.181802} {\bibfield  {journal} {\bibinfo
  {journal} {Phys. Rev. Lett.}\ }\textbf {\bibinfo {volume} {123}},\ \bibinfo
  {pages} {181802} (\bibinfo {year} {2019})},\ \Eprint
  {http://arxiv.org/abs/1907.12628} {arXiv:1907.12628 [astro-ph.CO]}
  \BibitemShut {NoStop}%
\bibitem [{\citenamefont {Chang}\ \emph {et~al.}(2021)\citenamefont {Chang},
  \citenamefont {Essig},\ and\ \citenamefont {Reinert}}]{Chang:2019xva}%
  \BibitemOpen
  \bibfield  {author} {\bibinfo {author} {\bibfnamefont {Jae~Hyeok}\
  \bibnamefont {Chang}}, \bibinfo {author} {\bibfnamefont {Rouven}\
  \bibnamefont {Essig}}, \ and\ \bibinfo {author} {\bibfnamefont {Annika}\
  \bibnamefont {Reinert}},\ }\bibfield  {title} {\enquote {\bibinfo {title}
  {{Light(ly)-coupled Dark Matter in the keV Range: Freeze-In and
  Constraints}},}\ }\href {\doibase 10.1007/JHEP03(2021)141} {\bibfield
  {journal} {\bibinfo  {journal} {JHEP}\ }\textbf {\bibinfo {volume} {03}},\
  \bibinfo {pages} {141} (\bibinfo {year} {2021})},\ \Eprint
  {http://arxiv.org/abs/1911.03389} {arXiv:1911.03389 [hep-ph]} \BibitemShut
  {NoStop}%
\bibitem [{\citenamefont {Blanco}\ \emph {et~al.}(2020)\citenamefont {Blanco},
  \citenamefont {Collar}, \citenamefont {Kahn},\ and\ \citenamefont
  {Lillard}}]{Blanco:2019lrf}%
  \BibitemOpen
  \bibfield  {author} {\bibinfo {author} {\bibfnamefont {Carlos}\ \bibnamefont
  {Blanco}}, \bibinfo {author} {\bibfnamefont {J.~I.}\ \bibnamefont {Collar}},
  \bibinfo {author} {\bibfnamefont {Yonatan}\ \bibnamefont {Kahn}}, \ and\
  \bibinfo {author} {\bibfnamefont {Benjamin}\ \bibnamefont {Lillard}},\
  }\bibfield  {title} {\enquote {\bibinfo {title} {{Dark Matter-Electron
  Scattering from Aromatic Organic Targets}},}\ }\href {\doibase
  10.1103/PhysRevD.101.056001} {\bibfield  {journal} {\bibinfo  {journal}
  {Phys. Rev. D}\ }\textbf {\bibinfo {volume} {101}},\ \bibinfo {pages}
  {056001} (\bibinfo {year} {2020})},\ \Eprint
  {http://arxiv.org/abs/1912.02822} {arXiv:1912.02822 [hep-ph]} \BibitemShut
  {NoStop}%
\bibitem [{\citenamefont {Blanco}\ \emph {et~al.}(2021)\citenamefont {Blanco},
  \citenamefont {Kahn}, \citenamefont {Lillard},\ and\ \citenamefont
  {McDermott}}]{Blanco:2021hlm}%
  \BibitemOpen
  \bibfield  {author} {\bibinfo {author} {\bibfnamefont {Carlos}\ \bibnamefont
  {Blanco}}, \bibinfo {author} {\bibfnamefont {Yonatan}\ \bibnamefont {Kahn}},
  \bibinfo {author} {\bibfnamefont {Benjamin}\ \bibnamefont {Lillard}}, \ and\
  \bibinfo {author} {\bibfnamefont {Samuel~D.}\ \bibnamefont {McDermott}},\
  }\bibfield  {title} {\enquote {\bibinfo {title} {{Dark Matter Daily
  Modulation With Anisotropic Organic Crystals}},}\ }\href {\doibase
  10.1103/PhysRevD.104.036011} {\bibfield  {journal} {\bibinfo  {journal}
  {Phys. Rev. D}\ }\textbf {\bibinfo {volume} {104}},\ \bibinfo {pages}
  {036011} (\bibinfo {year} {2021})},\ \Eprint
  {http://arxiv.org/abs/2103.08601} {arXiv:2103.08601 [hep-ph]} \BibitemShut
  {NoStop}%
\bibitem [{\citenamefont {Capparelli}\ \emph {et~al.}(2015)\citenamefont
  {Capparelli} \emph {et~al.}}]{Capparelli:2014lua}%
  \BibitemOpen
  \bibfield  {author} {\bibinfo {author} {\bibfnamefont {Ludovico~M.}\
  \bibnamefont {Capparelli}} \emph {et~al.},\ }\bibfield  {title} {\enquote
  {\bibinfo {title} {{Directional Dark Matter Searches with Carbon
  Nanotubes}},}\ }\href {\doibase 10.1016/j.dark.2015.08.002} {\bibfield
  {journal} {\bibinfo  {journal} {Phys. Dark Univ.}\ }\textbf {\bibinfo
  {volume} {9-10}},\ \bibinfo {pages} {24--30} (\bibinfo {year} {2015})},\
  \bibinfo {note} {[Erratum: Phys.Dark Univ. 11, 79--80 (2016)]},\ \Eprint
  {http://arxiv.org/abs/1412.8213} {arXiv:1412.8213 [physics.ins-det]}
  \BibitemShut {NoStop}%
\bibitem [{\citenamefont {Romao}\ \emph {et~al.}(2023)\citenamefont {Romao},
  \citenamefont {Catena}, \citenamefont {Spaldin},\ and\ \citenamefont
  {Matas}}]{sdata}%
  \BibitemOpen
  \bibfield  {author} {\bibinfo {author} {\bibfnamefont {Carl~P.}\ \bibnamefont
  {Romao}}, \bibinfo {author} {\bibfnamefont {Riccardo}\ \bibnamefont
  {Catena}}, \bibinfo {author} {\bibfnamefont {Nicola~A.}\ \bibnamefont
  {Spaldin}}, \ and\ \bibinfo {author} {\bibfnamefont {Marek}\ \bibnamefont
  {Matas}},\ }\href {\doibase 10.5281/zenodo.7615091} {\enquote {\bibinfo
  {title} {Chiral phonons as dark matter detectors -- supplementary data},}\ }
  (\bibinfo {year} {2023})\BibitemShut {NoStop}%
\end{thebibliography}%

\end{document}